\documentclass[10pt,journal,compsoc]{IEEEtran}
\usepackage{graphicx}
\usepackage{textcomp}
\usepackage{wrapfig}
\usepackage[export]{adjustbox}
\usepackage[table,xcdraw]{xcolor}
\usepackage{multirow}
\usepackage{dblfloatfix}
\usepackage[ruled,linesnumbered]{algorithm2e}
\usepackage[misc,geometry]{ifsym}
\ifCLASSOPTIONcompsoc
  \usepackage[nocompress]{cite}
\else
  \usepackage{cite}
\fi

\ifCLASSINFOpdf
  
\else
  
\fi

\usepackage{amsmath}

\usepackage{algorithmic}

\usepackage{array}

\ifCLASSOPTIONcompsoc
 \usepackage[caption=false,font=footnotesize,labelfont=sf,textfont=sf]{subfig}
\else
 \usepackage[caption=false,font=footnotesize]{subfig}
\fi

\usepackage{fixltx2e}

\usepackage{multirow}

\usepackage{booktabs}
\usepackage{hhline}
\DeclareUnicodeCharacter{2212}{-}
\usepackage{xcolor}
\usepackage{caption}
\usepackage{booktabs}
\usepackage{multirow}
\usepackage{amssymb}
\usepackage{pifont}
\newcommand{\cmark}{\ding{51}}%
\newtheorem{definition}{Definition}
\DeclareMathAlphabet{\mathpzc}{OT1}{pzc}{m}{it}
\usepackage[]{svg}
\newcommand{\lIfElse}[3]{\lIf{#1}{#2 \textbf{else}~#3}}

\begin{document}
\title{Adversarial Analysis of the Differentially-Private Federated Learning in Cyber-Physical Critical Infrastructures}

\author{Md Tamjid Hossain,
        Shahriar Badsha,
        Hung La\textsuperscript{\Letter},
        Haoting Shen,
        Shafkat Islam,
        Ibrahim Khalil,
        and Xun Yi
\IEEEcompsocitemizethanks{\IEEEcompsocthanksitem Md Tamjid Hossain and Hung La are with the ARA lab in the Computer Science and Engineering Dept. of the University of Nevada, Reno, USA.
\IEEEcompsocthanksitem Shahriar Badsha is with Bosch Engineering, North America.
\IEEEcompsocthanksitem Haoting Shen is with the School of Cyber Science and Technology, Zhejiang University, China.
\IEEEcompsocthanksitem Shafkat Islam is with the Dept. of Computer Science at Purdue University.
\IEEEcompsocthanksitem Ibrahim Khalil and Xun Yi are with the School of Science, RMIT University, Melbourne, VIC 3000, Australia.
\IEEEcompsocthanksitem(\Letter) Hung La is the corresponding author. Email: hla@unr.edu
}
}


\IEEEtitleabstractindextext{%
\begin{abstract}
Federated Learning (FL) has become increasingly popular to perform data-driven analysis in cyber-physical critical infrastructures. Since the FL process may involve the client's confidential information, Differential Privacy (DP) has been proposed lately to secure it from adversarial inference. However, we find that while DP greatly alleviates the privacy concerns, the additional DP-noise opens a new threat for model poisoning in FL. Nonetheless, very little effort has been made in the literature to investigate this adversarial exploitation of the DP-noise. To overcome this gap, in this paper, we present a novel adaptive model poisoning technique {\tt $\alpha$-MPELM} through which an attacker can exploit the additional DP-noise to evade the state-of-the-art anomaly detection techniques and prevent optimal convergence of the FL model. We evaluate our proposed attack on the state-of-the-art anomaly detection approaches in terms of detection accuracy and validation loss. The main significance of our proposed {\tt $\alpha$-MPELM} attack is that it reduces the state-of-the-art anomaly detection accuracy by $6.8\%$ for {\tt norm} detection, $12.6\%$ for {\tt accuracy} detection, and $13.8\%$ for {\tt mix} detection. Furthermore, we propose a Reinforcement Learning-based DP level selection process to defend {\tt $\alpha$-MPELM} attack. The experimental results confirm that our defense mechanism converges to an optimal privacy policy without human maneuver. 
\end{abstract}

}

\IEEEoverridecommandlockouts
\IEEEpubid{\makebox[\linewidth]{*This work has been submitted to the IEEE for possible publication. Copyright may be transferred without notice, after which this version may no longer be accessible.} }
\maketitle
\IEEEpubidadjcol
\IEEEdisplaynontitleabstractindextext

\IEEEpeerreviewmaketitle

\IEEEraisesectionheading{\section{Introduction}\label{sec:introduction}}

\IEEEPARstart{C}{yber-physical} critical infrastructures (CPCIs) comprises the essential assets equipped with the latest cyber-physical system (CPS) components for the safe and smart functioning of modern society and economy. However, despite the preventive measures and privacy awareness, a series of malicious attacks and campaigns have taken place targeting CPCIs (e.g., smart grids, water treatment plants, etc.) over the last decade. Particularly, the energy sector is a potential target for cyberattackers since it enables all other CPCI sectors \cite{CISA_2020}. Disrupting the operation of the energy sector, specially the smart grid, through cyberattacks can bring catastrophic consequences to other CPCIs. Therefore, smart grids get frequent attention from cyberattackers (e.g., Dragonfly [ICS-ALERT-14-176-02A], Stuxnet \cite{case2016analysis}).

A crucial feature of many cyberattacks on CPCIs is harnessing and modifying the sensitive sensor readings through available system vulnerabilities and backdoors. For example, unauthorized access to the smart meter readings can put the client's privacy at stake. In addition, tampering with those meter readings can mislead the grid controller to put the system into an unsafe operating condition (e.g., incorrect state estimation, wrong energy consumption prediction, asynchronism of the generators, etc.). Similarly, Machine Learning (ML) techniques including Federated Learning (FL)\cite{mcmahan2017communication} that are, nowadays, used for numerous industrial operations in CPCIs can also posses a threat of data privacy leakage of the clients \cite{wen2021feddetect,su2021secure,taik2020electrical}.\\
\indent To mitigate this, a significant body of research including data-driven privacy-preservation methods \cite{li2020privacy,mo2019efficient,zheng2022aggregation} has been carried out lately. Among the current privacy-preservation techniques, Differential Privacy (DP) \cite{dwork2006calibrating} has emerged as a standard privacy specification in the last few years due to its efficacy in preserving the clients' confidential information. It mostly utilizes a randomized noise-adding mechanism following well-known statistical distributions (e.g., Laplace, Gaussian, Exponential, etc.). Moreover, due to the provable privacy guarantee (in terms of \textit{privacy loss} measure) and the low computational cost, DP is proposed to be applied in various FL-based applications across CPCIs \cite{abadi2016deep, geyer2017differentially,zhao2020local,zhou2022differentially}. Nevertheless, as we show later in this paper, an intelligent attacker can leverage the additional noise (introduced to ensure DP) to conduct poisoning attacks in FL. Therefore, it is non-trivial to perform the adversarial analysis of the DP mechanism in the clients' confidential data-driven FL applications across CPCIs.

\subsection{Motivations}
\label{Motivations}
Though a good number of research are carried out on how DP can protect the privacy of the users \cite{dwork2006calibrating,abadi2016deep, geyer2017differentially,zhao2020local,hu2020personalized,zhou2022differentially,wen2021feddetect,li2022multi,sun2020ldp,truex2020ldp}, a limited effort is given on how it can be exploited to conduct security and integrity attacks. Specifically, only a few recent research \cite{giraldo2017security_2,giraldo2020adversarial,hossain2021PSU,hossain2021DeSMP} on the privacy and security challenges in the CPCI domain points out the exploitation opportunity of DP. Particularly, in FL-based CPCIs \cite{taik2020electrical,su2021secure,wen2021feddetect,zhou2022differentially}, DP can introduce new attack vectors for conducting stealthy poisoning attacks \cite{hossain2021DeSMP}. For instance, the attacker can craft malicious noise in the form of DP-noise and inject it into a few compromised model parameters. Now, if the magnitude of that adversarial noise is not too large (yet large enough to damage the system's performance), it would be difficult for the anomaly detector to identify and flag this as malicious from benign DP-noise. 

To conduct such DP-exploited attacks, the challenge of the attacker is \textit{to maintain the attack stealthiness while maximizing the attack impact}. To maintain the attack stealthiness (or to evade the anomaly detectors), the malicious noise must possess similar statistical behavior as the DP-noise. For this, the adversary needs to draw the false noise from an adversarial distribution that holds similar properties as a benign statistical distribution (e.g., Gaussian). Then, due to the differential structure of the malicious noise, the anomaly detector would be misguided to flag this noise as non-malicious. On the other hand, to maintain persistent damage, the noise injection process must follow any adaptive poisoning technique. That means the attacker needs to dynamically adjust the attack magnitude with the learning progression. We aim to find out the impact of such poisonings on the FL processes under adversarial presence. 

Fig. \ref{fig:LDP-FL} illustrates the attack vectors in the context of a Differentially Private Federated Learning, hereinafter referred to as the `{\tt DPFL}' process. In the case of a data poisoning attack, the attacker injects the adversarial noise into the differentially private training data whereas, in a model poisoning attack, the malicious noise is added into the differentially private model parameters. We mainly focus on the model poisoning case since it is more powerful, yet harder to conduct than the data poisoning. The adversarial process is further explained in section \ref{BasicAttack}.


\textbf{Research Gap.}
\label{ResearchGap} While several model poisoning attacks and defense models have been proposed in the literature in this direction, several limitations are observed. For example, Byzantine-robust aggregation \cite{fang2020local}, \textit{Krum} \cite{blanchard2017machine}, \textit{Bulyan} \cite{guerraoui2018hidden}, \textit{Trimmed Mean} \cite{yin2018byzantine}, \textit{Median} \cite{yin2018byzantine} schemes are proposed to defend poisoning attacks that exploits the vulnerabilities of FL aggregation rules. \textit{Nonetheless, they neither consider maintaining the attack stealthiness nor analyzes the opportunity of DP-exploitation in their attack methods}.

A novel DP-exploited false data injection attack (FDIA) is proposed in \cite{giraldo2020adversarial} which evades a classifier meant to detect anomalies. Additionally, they develop a log-likelihood ratio test-based anomaly detector as a countermeasure to defend against such DP-exploited FDIAs. \textit{Nevertheless, how these DP-exploited attacks perform in the federated and multi-agent settings is unclear from their studies. Also, how the attacker adjusts the degree of poisoning is missing}.

To seek how DP can be exploited in the context of FL, we formulate a global DP (GDP) exploited stealthy model poisoning attack earlier in \cite{hossain2021DeSMP}. \textit{However, how these attacks can achieve persistence in a Local Differentially private Federated Learning, hereinafter referred to `{\tt $\mathcal{L}$-DPFL}' environment remains as a question}. We analyze and subsequently answer these research questions through comparative analysis and empirical evidence in this paper.

\textbf{Our Work.}
\label{OurWork} with works mentioned above, in our scheme, we perform the first systematic study on the exploitation of the differential Gaussian noise to craft stealthy local model poisoning attacks in the FL domain. In particular, \textit{we devise a persistent and stealthy model poisoning attack that exploits the local DP (LDP) technique and exploit it to evade the state-of-the-art anomaly detectors while impairing the FL model utility}.

\subsection{Main Contribution}
\label{MainContributions}
Our key contributions can be summarized as follows:
\begin{figure}[!t]
    \centerline{\includegraphics[width=\columnwidth]{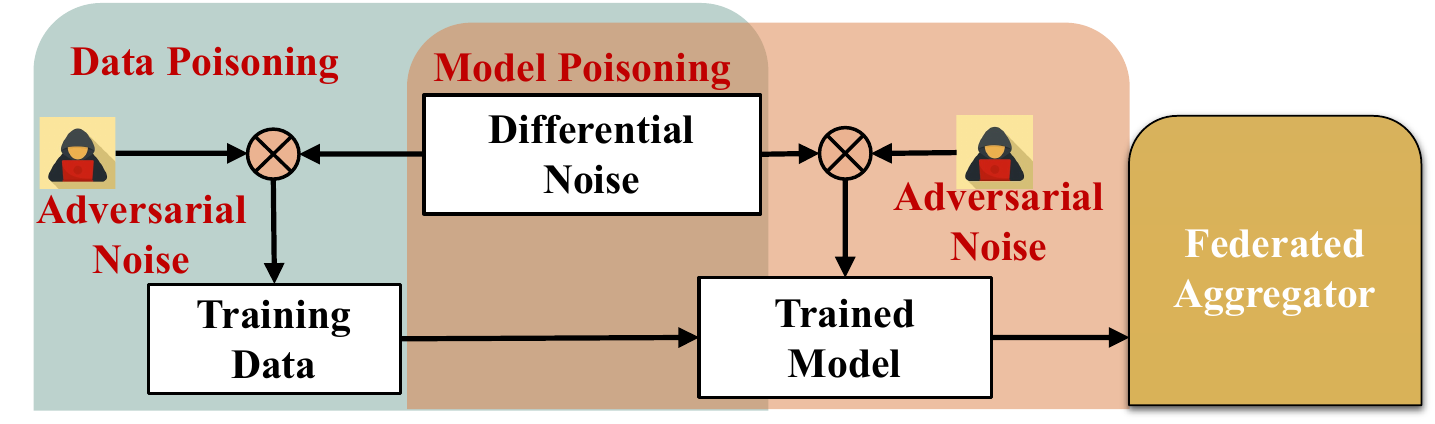}}
    \caption{DP-exploited data and model poisoning attacks in FL}
    \label{fig:LDP-FL}
\end{figure} 
\begin{itemize}
    \item We perform the first study on the exploitation of the differential Gaussian noise to conduct stealthy local model poisoning attacks in FL-based CPCIs.
    \item We perform our proposed attack over a smart grid metering dataset. 
    To maintain the attack stealthiness, and persistence, we propose an adaptive model poisoning technique ({\tt $\alpha$-MPELM}). We evaluate our novel attack on state-of-the-art anomaly detectors and show that our proposed attack can significantly evade these detection techniques.

    \item We propose to limit the attack surface by intelligently selecting the DP-noise level of the nodes through \textit{Reinforcement Learning (RL)}. We call it a Reinforcement Learning-assisted Differential Privacy level selection ({\tt rDP}) algorithm. Our evaluation shows that the proposed {\tt rDP}-algorithm converges to an optimal policy, thus disincentivizing the adversarial motivations.
    
\end{itemize}

\textbf{Roadmap.}
\label{Roadmap}
In section \ref{sec:Preliminaries}, we discuss the preliminaries. Section \ref{sec:LitReview} outlines some contrasting points between this work and state-of-the-art literature. Section \ref{sec:probFormulation} outlines the threat model and covers the basic mechanism of a DP-exploited poisoning attack. Our main contributions related to the attack-defense strategies are presented in section \ref{sec:OurAttack} and \ref{sec:rDPmodelling}. Section \ref{sec:expAnalysis} provides empirical evaluations of our proposed attack-defense strategies. We conclude the paper with some future research directions in section \ref{sec:conclusionAndFutureWorks}.

\textbf{Notation \& Keywords.}
\label{NotationKeywords} Table \ref{symbolTable} describes the major symbols used in this paper. We use `smart meters', `clients', `edge nodes', and `nodes' interchangeably throughout the rest of the article. Also, `aggregator' \& `remote station' have been used interchangeably.

\section{Preliminaries}
\label{sec:Preliminaries}

In CPCIs, the sensory data hold the private and confidential information of the clients and organizations. Oftentimes, the CPCI authorities collect and store sensory data in a central server which are then used for improving the performance of their ML-based applications and optimizing the operational states. For example, an electric utility company can utilize the energy consumption data to model their load balancing schemes and future demands. Nonetheless, an adversary can divulge clients' confidential information, for example, their whereabouts and the energy usage pattern from the ML training data. Sensitive information regarding the operational states of any CPCI can also be inferred and successively modified through the manipulation of model parameters (i.e., weights and biases) \cite{hao2019efficient,su2021secure,yin2021comprehensive}.
\begin{table*}[t!]
\centering
\caption{List of major symbols and their description}
\label{symbolTable}
 \begin{adjustbox}{max width=\linewidth}
\begin{tabular}{||c l c l c l c l||} 
 \hline
 Symbols & Description & Symbols & Description & Symbols & Description & Symbols & Description \\ 
 \hline\hline
$f_a$ & Adversarial distribution & $\mathcal{N}_a$ & Adversarial noise profile & $\eta_a$ & Adversarial noise & $\mu_a$ & Attack impact\\

$A$ & Action space & $R$ & Average global reward & $\beta_2$ & Accuracy benchmark & $f_0$ & Benign Gauss. distribution\\

$\eta_b$ & Benign noise & $\psi$ & Balancing param. & $\mathcal{C}$ & Clipping threshold & $\xi$ & Clipping technique\\

$\mathcal{R}$ & Detection range & $\tau$ & Detection threshold & $\upsilon$ & Deviation of model updates & $\gamma$ & Degree of poisoning\\

$\zeta$ & Discount factor & $k$ & Edge node & $\Delta w_e$ & Expected update & $\Delta w_g$ & Global update\\

$\eta$ & Gaussian DP-noise & $\rho$ & Factor of proportionality &  $f_l$ & Historical federated loss & $\alpha$ & Learning rate\\

$\mathpzc{D}$ &Local training data & $k_1$ & Lagrange multiplier & $\mathpzc{V}$ & Local validation set & $\ell$ & Loss function\\

$\mathpzc{R}$ & Loss ratio & $\Delta w$ & Model update & $\mathpzc{J}$ & Mini batch of local data & $\theta$ & Mean\\

$\hat{\eta}$ & Max DP-noise & $\tau^{'}$ & Modified detection threshold & $\mathpzc{W}$ & Max $\ell_2$-norm & $\beta_1$ & Norm detection benchmark\\

$b$ & No. of benign models & $a$ & No of malicious models & $\Phi$ & Objective function & $\delta$ & Probability of privacy leakage\\

$\Pi$ & Privacy accountant & $\varepsilon$ & Privacy loss & $Q$ & Q-table & $\Delta Q^{*}$ & Q-table converged \\

$t$ & episode & $\Delta w_r$ & Received update & $r$ & Reward & $\beta$ & Reward function \\

$N$ & Set of total nodes & $\mathcal{S}$ & Sensitivity & $\Delta w_m$ & Set of malicious updates & $M$ & Set of compromised nodes \\

$B$ & Set of benign nodes & $\xi$ & Set of noisy clipped updates & $S$ & State space & $m_l$ & Set of attacker's loss \\

$s$ & State & $d$ & Square of $L_2$ distance & $\mathcal{K}$ & Total available nodes & $t$ & Total participating nodes \\

$m$ & Total compromised nodes & $T$ & Total communication episodes & $\sigma^2$ & Variance & $\mathcal{L}$ & Validation loss\\
[1ex]
 \hline
\end{tabular}
 \end{adjustbox}
\end{table*}
\subsection{Federated Learning (FL) Mechanism}
\label{FLMechanism}
To prevent data leakage in ML, McMahan et al. propose the FL process \cite{mcmahan2017communication} following a multi-node environment. In FL, the model is trained on a diffuse network of edge nodes using local data. Therefore, the client’s private data do not leave the edge nodes. This ensures a level of privacy guarantee. 
Particularly, the FL algorithm performs the following three steps in each episode.

\textbf{Step I.} The global aggregator sends the global model parameters to all participating edge nodes. 

\textbf{Step II.} The nodes train their local models utilizing their local training data and the received global model parameters. To update the local model parameters, the nodes can use any optimization algorithm (e.g., batch gradient descent (BGD), stochastic gradient descent (SGD), mini batch gradient descent (mBGD), etc.) though mBGD is the most balanced one in this context. For instance, following mBGD, the $k$th node ($k \in N$) performs local update $\Delta w^{(t)}_k$ at episode $t = \{1, 2, ..., T\}$ as follows.
\begin{equation}
    \Delta w^{(t)}_k = w^{(t)}_k - w^{(t)}_g
    \label{eqn:lmpu}
\end{equation}
where $w^{(t)}_g$ is the global model and $w^{(t)}_k$ is the optimized local model. Here, $w^{(t)}_k$ is computed by taking a step towards the mini batch gradient descent as follows:
\begin{equation}
    w^{(t)}_k \leftarrow w^{(t)}_g - \alpha . \frac{\partial \Phi(w^{(t)}_g, \mathpzc{J}_k^{(t)})}{\partial w^{(t)}_g}
    \label{eqn:mBGD-lmpu}
\end{equation}
where $\alpha$ is the learning rate, $w^{(t)}_g$ is the global model parameters, $\mathpzc{J}_k^{(t)}$ is the mini batches of the local training data $\mathpzc{D}_k$ and $\Phi(w^{(t)}_g, \mathpzc{J}_k^{(t)})$ is the objective function to be minimized. The local models are then sent to the global aggregator.

\textbf{Step III.}
The global aggregator aggregates all the trained local models using any state-of-the-art aggregation rules (e.g., {\tt FedeAvg} \cite{mcmahan2017communication}, {\tt FedSGD} \cite{chen2016revisiting}, Byzantine-robust aggregation \cite{fang2020local}, \textit{Krum} \cite{blanchard2017machine}, \textit{Trimmed Mean} \cite{yin2018byzantine}, \textit{Median} \cite{yin2018byzantine}, etc.). Formally, according to the naive mean aggregation rules \cite{mcmahan2017communication}, the global model update, $w^{(t+1)}_{g}$ at the end of episode $t$ would be 
\begin{equation}
    w^{(t+1)}_{g} \leftarrow w^{(t)}_g +  \frac{1}{n}\left[(\sum_{k=1}^{n} \Delta w^{(t)}_k \right]
    \label{eqn:gmpu}
\end{equation}
Nonetheless, the naive mean aggregation rule \cite{mcmahan2017communication} is non-robust under an adversarial setting since the attacker can easily manipulate the global model using only one edge node \cite{yin2018byzantine}. Therefore, in our setting, we will utilize a more realistic averaging-based aggregation approach (detailed in section \ref{architecture}). Towards this direction, some state-of-the-arts suggest adopting stochastic client selection processes based on effective participation and fairness to achieve faster convergence and better FL accuracy \cite{huang2022stochastic}. However, since our primary focus in this paper is 
to analyze the adversarial impacts on the privacy-enhancing FL processes, we follow the general random sampling methods \cite{bressert2012scipy} for simplicity.

\subsection{Local Differential Privacy (LDP) with Gaussian Noise}
\label{LDPwithNoise}
\noindent Despite the inherent data privacy protection, FL is found to be vulnerable to membership inference attacks (MIAs) \cite{geyer2017differentially}. To address this vulnerability and protect the client's confidentiality, the DP-technique is proposed to be incorporated in various FL-based applications \cite{zhou2022differentially,geyer2017differentially}. Typically, two approaches are followed to achieve DP: (1) global DP (GDP) \cite{abadi2016deep, wei2020federated}, and (2) local DP (LDP) \cite{sun2020ldp,truex2020ldp}. While GDP seeks to perturb models in the aggregation phase, LDP perturbs models before sending those out from the edge nodes to the remote station. Since LDP provides a stricter definition of privacy, it has become a popular choice to preserve clients' privacy in FL. We also follow LDP in this paper.

Nonetheless, many variants of both GDP and LDP techniques can be found in the literature today which are developed to cater to varying conditions and requirements \cite{arachchige2019local, zhou2022differentially, geyer2017differentially, wei2020federated, truex2020ldp, zhao2020local, wen2021feddetect, hu2020personalized}. Although these variants follow several DP-mechanisms and their modifications including randomized response (satisfies $\varepsilon$-DP), Laplace (satisfies $\varepsilon$-DP), Gaussian (satisfies $(\varepsilon,\delta)$-DP), Exponential (satisfies $\varepsilon$-DP), Geometric (satisfies $\varepsilon$-DP) and Binomial (satisfies $(\varepsilon,\delta)$-DP) mechanisms, in general, they follow the same principle which is adding randomized noise or responses to the original data to protect the sensitive information of the clients \cite{el2022differential}. For example, \cite{geyer2017differentially, zhou2022differentially, wei2020federated, hu2020personalized} follows Gaussian mechanism to develop their `{\tt DPFL}' processes whereas \cite{wen2021feddetect, zhao2020local,arachchige2019local} adopts randomized response mechanism. Likewise, \cite{truex2020ldp} propose {\tt LDP-Fed} which relay on the Exponential mechanism and \cite{sun2020ldp} follow the Geometric mechanism in their proposed {\tt LDP-FL} method. \textit{Since, the Gaussian mechanism has the following two advantages over other DP mechanisms and thus, one of the most widely adopted techniques, we consider it to achieve a relaxation of pure DP which is $(\varepsilon,\delta)$-DP in this paper}.
\begin{itemize}
    \item \textit{Additive noise.} The sum of two jointly or independent Gaussian distributions ($X$ and $Y$) is a new Gaussian distribution ($aX+bY$; $a$ and $b$ are constants); therefore it is easier to analyze it statistically
    \item \textit{Natural noise.} The added Gaussian noise holds similar statistical properties as the natural noise that might appear in a query result from a database
\end{itemize}
Usually, in the Gaussian mechanism, the noise $\eta$ is drawn from a zero-mean Gaussian distribution having a probability density function (PDF) as
\begin{equation}
    f_0(r) = \frac{1}{\sqrt{2\pi}\sigma_r}e^{-\frac{(r-\theta)^2}{2\sigma_r^2}}
    \label{eqn:benignDist}
\end{equation} where $\theta$ is the mean and $\sigma^2$ is the variance. However, $(\varepsilon,\delta)$-DP only satisfies if $\sigma \geq c\mathcal{S}/\varepsilon$ where $c^2 > 2\ln(1.25/\delta)$. Here, $\varepsilon$ is the permitted privacy loss or simply, privacy budget, $\delta$ is the probability of exceeding the privacy budget, and $\mathcal{S}$ is the local sensitivity. This can be formally defined as \cite{duchi2013local}:

\begin{definition}
\label{definition}
Let $\mathcal{X}$ be a set of possible values and $\mathcal{Y}$ the set of noisy values. A local randomizer $\mathpzc{M}$ is $(\varepsilon,\delta)$-locally differentially private (LDP)
if $\forall x, x' \in \mathcal{X}$ and $\forall y \in \mathcal{Y}$:
$Pr\left[\mathpzc{M}(x) = y\right] \leq e^\varepsilon \times Pr\left[\mathpzc{M}(x')=y\right]+\delta$
\end{definition}
Under the aforementioned constraint and definition, in {\tt $\mathcal{L}$-DPFL}, it is important to keep track of the spent privacy budget to protect privacy in case of multiple new queries. Since $\delta$ is accumulative and grows with the consecutive queries, a moments accountant technique similar to \cite{abadi2016deep} could be useful to keep track of it and stops the training once a predefined threshold is reached. Following {\tt $\mathcal{L}$-DPFL}, the final local model update of the $k$th edge node is 
\begin{equation}
    \widetilde{\Delta w}^{(t)}_k \leftarrow \Delta w^{(t)}_k + \eta
    \label{eqn:DP-lmpu}
\end{equation}
and, after episode $t$, the global model update is
\begin{equation}
  w^{(t+1)}_{g} \leftarrow w^{(t)}_g +  \frac{1}{n}\left[\sum_{k=1}^{n} \widetilde{\Delta w}^{(t)}_k\right]
  \label{eqn:DP-gmpu}
\end{equation}

\section{Literature Review}
\label{sec:LitReview}
In this section, we point out some significant contrasting contributions between this work and state-of-the-arts.

\subsection{Model Poisoning Attacks and Defenses in FL}
\label{LocalModelPoisoningAttack}
\noindent Lately, the ML community proposed a substantial number of model poisoning attack-defense methods which in principle, are also applicable for the state-of-the-art FL mechanisms \cite{fang2020local, blanchard2017machine,guerraoui2018hidden,yin2018byzantine,zhou2022differentially,awancontra,mothukuri2021survey,gao2021secure}. However, most of them focus on the \textit{Byzantine failures} of the participating edge nodes where a single or a group of curious/semi-honest/malicious nodes manipulate either the local raw data (data poisoning) or the model parameters (model poisoning) and send those to the global aggregator instead of the true updates. To limit the impact of the Byzantine failures, a Byzantine-tolerant distributed random gradient descent algorithm named \textit{Krum} has been proposed in \cite{blanchard2017machine}. \textit{Krum} follows a combination of \textit{majority-based} and \textit{squared-distance} approaches to compute the $n-f-2$ local models for each local model, $w_k$ where $n$ being the total participating models and $f$ being the Byzantine models. \textit{Although \textit{Krum} has the theoretical guarantees for the convergence when $f < (n-2)/2$, it is not suitable for CPCIs where a large number of edge nodes are jointly training a federated model. Likewise, other Byzantine-robust FL methods (e.g., \textit{Trimmed Mean} \cite{yin2018byzantine}, \textit{\textit{Bulyan}} \cite{guerraoui2018hidden}, \textit{Median} \cite{yin2018byzantine}), despite being successful against data poisoning attacks, performs poorly in defending the model poisoning attacks \cite{fang2020local}}.

Two defense techniques are proposed in \cite{fang2020local} against local model poisoning attacks in Byzantine-robust FL. The authors set the attacker's goal as \textit{directed deviation} and \textit{deviation}. Following the \textit{directed deviation} goal, the attacker aims to deviate a global model parameter the most towards the inverse of the before-attack direction. Under the \textit{deviation} goal, the direction change of the global model parameter is not considered. They show that \textit{Krum}, \textit{Trimmed Mean}, and \textit{Median} are vulnerable to their attack. As countermeasures they propose \textit{ERR} and \textit{LFR} techniques which are the generalization of two earlier data poisoning defense techniques: \textit{RONI} \cite{barreno2010security} and \textit{TRIM} \cite{jagielski2018manipulating}. \textit{ERR} removes the local models that have large negative impacts on the error rate of the global model whereas \textit{LFR} removes the local models having large losses. A mix detection technique using the union of \textit{ERR} and \textit{LFR} (i.e., \textit{ERR} + \textit{LFR}) is also proposed and found effective in certain cases. \textit{Nonetheless, their attack model aims to attain maximum degradation in model utility whereas attaining stealthiness through adaptive manipulation can be another crucial criterion for the attacker \cite{giraldo2020adversarial}. Moreover, they overlook the privacy issues, oftentimes required by the FL users}.

A recently published closely related work to ours \cite{zhou2022differentially} studies the problem of both data and model poisoning attacks in FL. Particularly, they propose a \textit{weight-based detection} method that can detect and filter malicious or anomalous model parameters in the intermediary testing phases using a validation dataset. Their \textit{weight-based detection} technique comprised of two separate detection mechanisms: {\tt norm} detection and {\tt accuracy} detection which are, interestingly, similar to the \textit{ERR} and \textit{LFR} detection \cite{fang2020local} in terms of underlying operational principle. Their {\tt mix} detection technique also follows the mix detection of \cite{fang2020local}. Nonetheless, unlike \cite{fang2020local}, they introduce $\gamma$ as a degree of influence for their {\tt mix} detection technique (\textit{remark:} we use the symbol $\gamma$ in this paper for describing the degree of poisoning which bears a different meaning than this). They evaluate their {\tt norm, accuracy,} and {\tt mix} detection approaches in the presence of randomized malicious devices (RMD). Furthermore, they introduce a multi-layer ($\varepsilon,\delta$)-GDP technique to balance the privacy-utility trade-off of DP effectively. Specifically, they apply DP to the end devices' training data and the edge nodes' and cloud servers' aggregated parameters. \textit{In contrast, to realize a stringent definition of privacy without loss of generality, we make use of LDP in this paper. Our edge node is analogous to the end devices of \cite{zhou2022differentially}. We focus on protecting the local model parameters than the raw training data. This is because the in-transit model parameters are more vulnerable to inference attacks than the raw training data. Also, the model poisoning attacks are more likely to cause irrevocable utility damages than the data poisoning attacks. Furthermore, while \cite{zhou2022differentially} considers DP-noise as a privacy-preservation tool, we consider DP-noise as an escape clause to conduct model poisoning in FL. Later, we show that our proposed attack can deceive their anomaly detection techniques more effectively than conventional RMD attacks}.

Similarly, to address the poisoning attacks on FL, \cite{li2021lomar} introduces a two-phase defense algorithm called \textit{LoMar}. \textit{LoMar} scores each local model update over the neighboring updates by measuring the relative distribution following kernel density estimation. Successively, it filters out the malicious models from the benign models. Nevertheless, their poisoning attacks are not conducted by leveraging the additional DP-noise. \textit{Contrarily, we consider the malicious noise for poisoning attack, to be drawn from an adversarial distribution having similar properties as any benign Gaussian distribution. Therefore, both the malicious and benign model updates would reflect similar statistical behaviors, which are difficult to be distinguished through the \textit{LoMar} technique as in \cite{li2021lomar}}. 
\begin{figure*}[!ht]
    \centerline{\includegraphics[width=\linewidth]{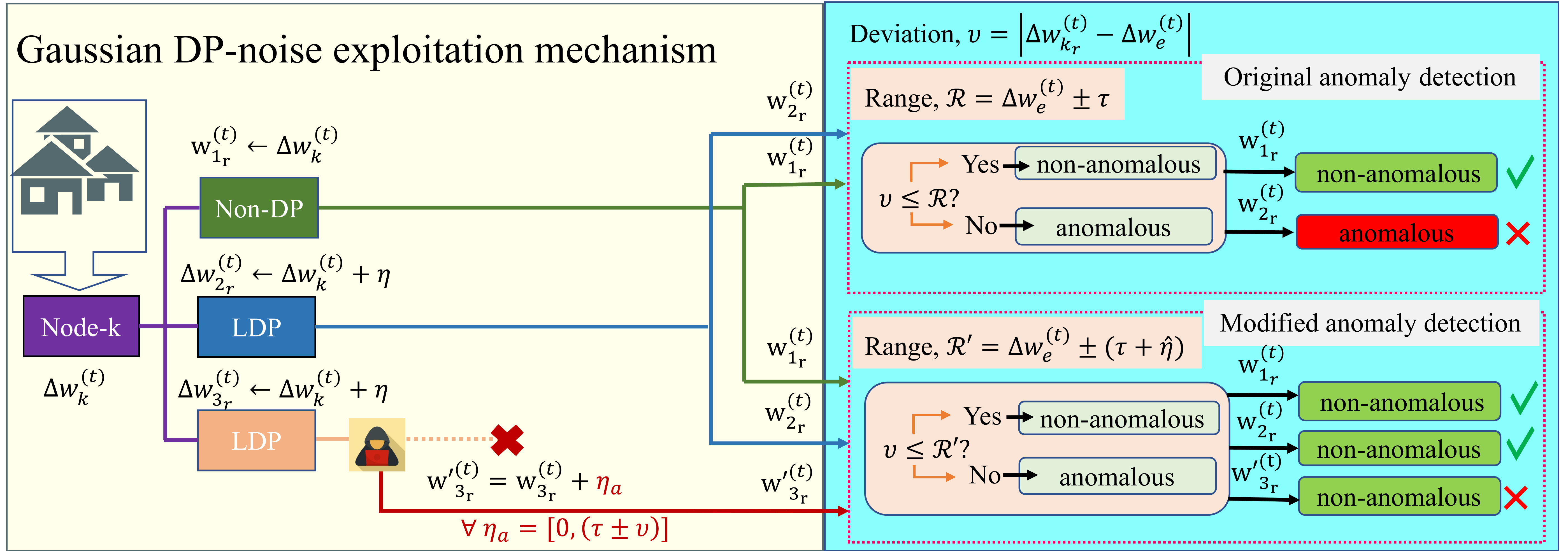}}
    \caption{Basic mechanism of Gaussian DP-noise exploitation}
    \label{fig:exploitationMechanism}
\end{figure*}

\subsection{Exploitation of DP and Countermeasures}
\label{DPexploitationCountermeasure}
\noindent Another line of research \cite{giraldo2017security_2, giraldo2020adversarial, hossain2021PSU}, though neither concentrates on the FL nor model poisoning attacks, yet very relevant to our research problem, studies the exploitation opportunities of DP in the realm of classification problems. Particularly, \cite{giraldo2020adversarial} studies the exploitation of DP-noise to degrade the system utility. They formulate an optimal adversarial distribution and draw adversarial noise from there. In addition, they propose a bad data detection (BDD)-based defense called \textit{DP-BDD}. Particularly, they model their optimal attack-defense following a game-theoretic approach (more exactly, a leader-follower sequential game) through which their \textit{DP-BDD} algorithm can be evaluated as a \textit{Nash equilibrium} point. textit{Although they emphasize maintaining the attack stealthiness, how their attack model performs in federated or any multi-agent settings is not clear. Also, how the attacker controls the degree of poisoning is missing. Furthermore, they do not consider limiting the attack surface in the first place, which we achieve through our {\tt rDP} algorithm in this paper}.

\begin{table}[!t]
\centering
\caption{Comparative analysis among {\tt DPFL} systems. Symbol: Addressed($\textbf{\cmark}$), Not addressed($\square$).
``\textbf{F}"ederated learning. ``\textbf{Po}"isoning attacks. ``\textbf{D}"ifferential Privacy (``\textbf{$\mathcal{L}$}"ocal DP or ``\textbf{$\mathcal{G}$}"lobal DP). ``\textbf{E}"xploitation of DP to conduct poisoning attacks. ``\textbf{T}"racking of privacy budget spending. ``\textbf{I}"ntelligent ``\textbf{P}"rivacy ``\textbf{L}"evel ``\textbf{S}"election strategy}
\resizebox{\linewidth}{!}{%
\begin{tabular}{lccccccc} 
\toprule
\multirow{2}{*}{\textbf{System }} & \multirow{2}{*}{\textbf{F }}                    & \multirow{2}{*}{\textbf{Po}}                   & \multicolumn{2}{c}{\textbf{D}}                                                                  & \multirow{2}{*}{\textbf{E}}                  & \multirow{2}{*}{\textbf{T}}                     & \multirow{2}{*}{\textbf{IPLS}}                                                                                                                    \\ 
\cline{4-5} &       &       &  \textbf{$\mathcal{L}$}    & \textbf{$\mathcal{G}$}    &           &   &    \\ 
\hline
Fang et al., 2020 \cite{fang2020local}            & \textbf{\cmark} & \textbf{\cmark} & $\square$                     & $\square$                     & $\square$                     & $\square$                     & $\square$                      \\
Giraldo et al., 2020 \cite{giraldo2020adversarial}         & $\square$ & \textbf{\cmark}  & \textbf{\cmark} & \textbf{\cmark} & \textbf{\cmark} & $\square$ & $\square$  \\
Zhao et al., 2020 \cite{zhao2020local}            & \textbf{\cmark} & $\square$  & \textbf{\cmark} & $\square$                     & $\square$ & \textbf{\cmark} & $\square$  \\
Hu et al., 2020 \cite{hu2020personalized}              & \textbf{\cmark} & $\square$ &  \textbf{\cmark} & $\square$                     & $\square$ & \textbf{\cmark} & \textbf{\cmark}  \\
Wen et al., 2021 \cite{wen2021feddetect}              & \textbf{\cmark} & $\square$ &  \textbf{\cmark} & $\square$                     & $\square$ & $\square$ & $\square$ \\
Zhou et al., 2022 \cite{zhou2022differentially}            & \textbf{\cmark} & \textbf{\cmark}  & $\square$ & \textbf{\cmark} & $\square$ & \textbf{\cmark} & \textbf{\cmark} \\
Li et al., 2022 \cite{li2022multi}              & \textbf{\cmark} & \textbf{\cmark} &  \textbf{\cmark} & $\square$                     & $\square$ & $\square$ & $\square$  \\
\textbf{This work}                & \textbf{\cmark} & \textbf{\cmark} &  \textbf{\cmark} & $\square$                     & \textbf{\cmark} & \textbf{\cmark} & \textbf{\cmark} \\
\bottomrule
\end{tabular}
}
\label{table:comparativeAnalysis}
\end{table}

Even though we address the Gaussian DP-noise exploitation in the context of GDP-based FL in \cite{hossain2021DeSMP}, a method has not been devised yet in the literature to keep the attack persistent, robust, and stealthy, especially in {\tt $\mathcal{L}$-DPFL}. If the attack is not persistent throughout the communication episodes, the random node selection method that almost every state-of-the-art FL processes adopt nowadays may cancel out any adversarial contribution. Moreover, if the attack is non-robust, the attack impact would be negligible. We show evidence of achieving attack stealthiness, persistence, and robustness in this paper. Table \ref{table:comparativeAnalysis} summarizes the comparison with the most related works discussed above.

\section{Problem Formulation}
\label{sec:probFormulation}

In this section, we present the basic mechanism of Gaussian noise exploitation and formulate the challenges in crafting an adversarial noise profile.

\subsection{Basic Mechanism of Gaussian Noise Exploitation}
\label{BasicAttack}

\textbf{DP not included.} Let us first consider a non-DP setting where the anomaly detector is expecting local model updates from all of the nodes as $\Delta w^{(t)}_{e}$. Let us also assume that instead of $\Delta w^{(t)}_{e}$, the detector receives local update $\Delta w^{(t)}_{k_r}$ from $k$th node. The detector raises an alarm if the received update, $\Delta w^{(t)}_{k_r}$ exceeds a pre-defined detection range, $\mathcal{R}$, i.e., $ \Delta w^{(t)}_{k_r} > \mathcal{R}$ where $\mathcal{R} = [\Delta w^{(t)}_{e} \pm \tau]$ and $\tau$ is the predefined detection threshold.

\noindent\textbf{DP included.} Now, consider that the authority enforces DP for the sake of privacy. Therefore, the received differentially private  local update with a maximum Gaussian noise $\pm \hat{\eta}$ would be $\widetilde{\Delta w}^{(t)}_{k_r} \leftarrow \Delta w^{(t)}_{k_r} \pm \hat{\eta}$. To avoid false-positive alarm for this non-malicious modified update, the detector needs to adjust its detection range as, $\mathcal{R}' = [\Delta w^{(t)}_{e} \pm \tau']$ where the new detection threshold, $\tau' = \pm(\tau+\hat{\eta})$. \textit{This adjustment in the detection range opens an additional (false) noise injection window for the attacker. The range is as follows:}

\begin{equation}
\begin{split}
Lower: \left[0, (\Delta w^{(t)}_{k_r} - \hat{\eta}) - (\Delta w^{(t)}_{e} - \tau')\right] \\ \Rightarrow \left[0, \Delta w^{(t)}_{k_r} - \Delta w^{(t)}_{e} + \tau\right] \Rightarrow \left[0, \tau - \upsilon \right]\\ 
Upper: \left[0, (\Delta w^{(t)}_{e} + \tau') - (\Delta w^{(t)}_{k_r} + \hat{\eta})\right] \\
 \Rightarrow \left[0, \Delta w^{(t)}_{e} - \Delta w^{(t)}_{k_r} + \tau\right]
 \Rightarrow \left[0, \tau + \upsilon \right]
\end{split}
\label{eqn:basicRange}
\end{equation}
where $\upsilon$ is the deviation of the local update from the expected update, i.e., $\upsilon = \Delta w^{(t)}_{k_r} - \Delta w^{(t)}_{e}$. The adversary can exploit this false noise injection or poisoning window (i.e., $[0, \tau \pm \upsilon]$) to craft an adversarial noise profile,
$\eta_a \leftarrow \mathcal{N}_a(\mu_a, \frac{\mathcal{S}}{\varepsilon})$ where $\mu_a$ is the desired deviated mean or simply attack impact. If noise is increased, deviation $\upsilon$ increases which in turn expands the poisoning window. \textit{\textit{In a nutshell, more privacy (i.e., large DP-noise) leads to more attack opportunities and more utility degradation.}}

Fig. \ref{fig:exploitationMechanism} illustrates this adversarial manipulation opportunity. When DP is not enforced, the original anomaly detector flags the update $w^{(t)}_{1_r}$ as non-anomalous if its' deviation $\upsilon$ lies within the detection range $\mathcal{R}$. However, when LDP is deployed, the additional Gaussian noise $\eta$, introduced to ensure LDP, may push $\upsilon$ beyond $\mathcal{R}$, leading to misclassification of the benign LDP-update $w^{(t)}_{2_r}$ to anomalous one. To cater $\eta$ and correctly classify $w^{(t)}_{2_r}$, the detector modifies its' detection range to $\mathcal{R'}$. Nevertheless, this slight modification in detection range can open poisoning window ($\left[0, (\tau \pm \upsilon)\right]$) for an strategic attacker. This may lead the detector to misclassify the malicious update $w'^{(t)}_{3_r}$ as non-anomalous. In short, the detector can not possibly distinguish between the benign and the malicious contribution if the adversarial noise magnitude does not exceed the poisoning window.

\subsection{Challenges in Crafting Adversarial Noise Profile}
\label{AdversarialUpdate}

One cannot assume that the attacker simply sets $\eta_a$ knowing the value of $\tau$ and $\upsilon$ since they are safeguarded with the anomaly detector. To understand the problem more clearly, let us consider that $i$ is a particular compromised edge node and $M$ is the set of all compromised nodes having cardinality of $m$ (i.e., $i \in M$ and $\lvert M \rvert = m$) in a {\tt $\mathcal{L}$-DPFL} setting. Then the number of benign edge nodes is $b = n - m$ where $n$ is the total number of participating edge nodes (i.e., $\lvert N \rvert = n$). Let us also consider $j$ is an individual benign edge node while the set of benign nodes is $B$ (i.e., $j \in B$ and $\lvert B \rvert = b$). Now, if the malicious noise is $\eta^{(t)}_{ai}$ and the benign DP-noise is $\eta^{(t)}_{bj}$, then $i$'s adversarial local update, $j$'s benign local update, and aggregated global update at episode $t$ can be represented by (\ref{eqn:M-DP-lmpu}), (\ref{eqn:B-DP-lmpu}), and (\ref{eqn:MB-DP-gmpu}) respectively.
\begin{equation}
    \widetilde{\Delta w}^{(t)}_i \leftarrow \Delta w^{(t)}_i + \eta^{(t)}_{ai}
    \label{eqn:M-DP-lmpu}
\end{equation}
\begin{equation}
    \widetilde{\Delta w}^{(t)}_j \leftarrow \Delta w^{(t)}_j + \eta^{(t)}_{bj}
    \label{eqn:B-DP-lmpu}
\end{equation}
\begin{equation}
    w^{(t+1)}_g \leftarrow w^{(t)}_g + \frac{1}{n}\left[\sum_{i=1}^{m} \widetilde{\Delta w}^{(t)}_i + \sum_{j=1}^{b} \widetilde{\Delta w}^{(t)}_j\right] \forall i\in M, j\in B
    \label{eqn:MB-DP-gmpu}
\end{equation}

However, here the main challenge for the attacker is to craft the adversarial noise profile, $\mathcal{N}_a(\mu_a, \frac{\mathcal{S}}{\varepsilon})$ and choose the magnitude of the adversarial noise, $\eta_a$ for subsequent FL episodes. As mentioned in section \ref{LocalModelPoisoningAttack}, this particular challenge is also addressed in \cite{fang2020local} where the authors attempt to solve it through a maximum utility degradation approach. Nonetheless, for maximum utility degradation, it is intuitive that a large amount of noise is required to inject into the model parameters continuously through subsequent communication episodes which may eventually lead to easier attack detection (i.e., violating the stealthiness goal of the attacker).

Another way to address this adversarial noise injection challenge is
to draw $\eta_a$ from an adversarial distribution, $f_a$ similar to a benign Gaussian distribution, $f_0$ and, then inject $\eta_a$ into total $m$ compromised local models. In other words, if $i$ is a compromised \textit{edge node} ($i \in N$- set of all participating nodes) out of the total $m$ compromised nodes, then the set of the malicious local model at episode $t$ is 
\begin{equation}
   \Delta w^{(t)}_m = \{\Delta w^{(t)}_i + \eta^{(t)}_{ai}\}_{i=1,2,...m}\;\forall i\in N \; \text{if}\;0 \leq m \leq n 
   \label{eqn:maliciousSet}
\end{equation}

Such optimal attack distribution, $f_a^*$ and the optimal attack impact, $\mu_a^*$ are derived and presented in \cite{giraldo2020adversarial} by solving a multi-criteria optimization problem that addresses two conflicting adversarial goals: \textit{(1) maximum damage,} and \textit{(2) minimum disclosure}. The goals are contradicting in nature from the adversarial point of view since \textit{maximum damage} can lead to easier attack detection whereas \textit{minimum disclosure} limits the damage. The optimal adversarial distribution, $f_a^*$, and the optimal attack impact, $\mu_a^*$ for the Gaussian mechanism are expressed as:
\begin{equation}
    f_a^*(x) = \frac{1}{\sqrt{2\pi}\sigma_x}e^{-\frac{(x-\theta-\sqrt{2\gamma}\sigma_x)^2}{2\sigma^2_x}} \;\text{and}\; \mu_a^* = \theta + \sqrt{2\gamma}\sigma_x
    \label{eqn:attackDist}
\end{equation}
where $\theta$ is the mean, $\sigma^2_x$ is the variance, and $\gamma$ is the stealthiness parameter (i.e., degree of poisoning). A high $\gamma$ implies the higher damage achieved. At the same time, it leads to a high probability of attack identification. Contrarily, if $\gamma = 0$, $f_a^* = f_0$ and $\mu_a^* = \theta$ (comparing (\ref{eqn:benignDist}) and (\ref{eqn:attackDist})) in which case, the attack impact is negligible. Therefore, the attacker needs to tune the degree of poisoning, $\gamma$ to an appropriate level for each FL episode so that the attack stealthiness can be achieved with satisfactory attack impact. Now, this raises below questions that we subsequently answer through the theoretical and empirical analysis of our proposed model in later sections (section \ref{adaptiveAttack}, \ref{rdpAlgodescription}, and \ref{AdversarialImpactAnalysis}) of this paper.
\begin{itemize}
    \item How does the attacker tune $\gamma$ at every FL episode?
    \item What are the attack impacts in {\tt $\mathcal{L}$-DPFL} based CPCIs?
    \item  What could be an effective defense against this attack that simultaneously protects the client's privacy and achieves satisfactory {\tt $\mathcal{L}$-DPFL} performance? 
\end{itemize}

\subsection{Threat Model}
\label{sec:threatModel}

\subsubsection{Attacker's Capability}
\label{attackersCapability}
We consider an attacker who returns false model parameters to an aggregator in a {\tt DPFL} system.
In the local DP setting, the attack can be launched either by compromising a few vulnerable edge nodes (insider threat) or their communication paths to the aggregator (outsider threat). As We are agnostic to the way in which the attacker is able to modify the model parameters, our proposed attack model covers both the insider and outsider threat. \textit{Particularly, we assume that the attacker gets unauthorized access to a few local models irrespective of the attack vector.} However, the number of compromised models should not be too large; otherwise, the global model can be manipulated without much effort and the attack would be very easy to conduct \cite{fang2020local}. We also assume that the attacker does not have significant knowledge about the benign participants. Moreover, the attacker cannot control the FL aggregation algorithm directly. Specifically, the attacker cannot change the local updates that are already on the aggregator's end. As in practice, the aggregator is equipped with high-security measures and is difficult to penetrate or compromise for an attacker. Therefore, we consider the aggregator as a non-compromised benign server.
\subsubsection{Attacker's Background Knowledge}
\label{attackersKnowledge}
In the insider threat model, the attacker gets unauthorized access to both the local training data and local model parameters of a few compromised nodes. On the other hand, in the outsider threat model, the attacker only gets unauthorized access to a few in-transit model parameters. \textit{Following a conservative approach, we assume the attacker's background knowledge is only limited to a few local model parameters for both insider and outsider threat models.} Notice that our proposed attack algorithm will perform much better if the attacker also gets access to the local training data.

Furthermore, following the state-of-the-arts \cite{smart2022understanding, dwork2019differential} on the importance of publishing the privacy budget ($\varepsilon$) value for gaining the trust of the clients, we consider that the attacker knows publicly available imposed $\varepsilon$ value and the noise distribution mechanism.

\subsubsection{Attacker's Goal}
\label{AttackersGoal}
Attacker's primary goal is to achieve \textit{(1) maximum damage} while \textit{(2) avoiding detection} in any stage of the attack. To achieve \textit{maximum damage}, the adversarial noise should be as large (or as small) as possible; nonetheless this large adversarial can be easily detected by any conventional anomaly detectors. On the other hand, for \textit{avoiding detection}, the adversarial noise should be as close to the upper (or lower) bound of the poisoning range given by (\ref{eqn:basicRange}); but that might not fulfill the first goal of the attacker. Therefore, the attacker needs to optimize these two conflicting goals to obtain \textit{optimal damage and stealthiness}.

\section{Proposed Model Poisoning Attack}
\label{sec:OurAttack}
In this section, we first describe the {\tt $\mathcal{L}$-DPFL} architecture that we use to develop our proposed attack model in conjunction with DP and FL. Then, we present our proposed adaptive model poisoning attack model.
\subsection{{\tt $\mathcal{L}$-DPFL} Architecture}
\label{architecture}
\begin{figure}[!t]
    \centerline{\includegraphics[width=\columnwidth]{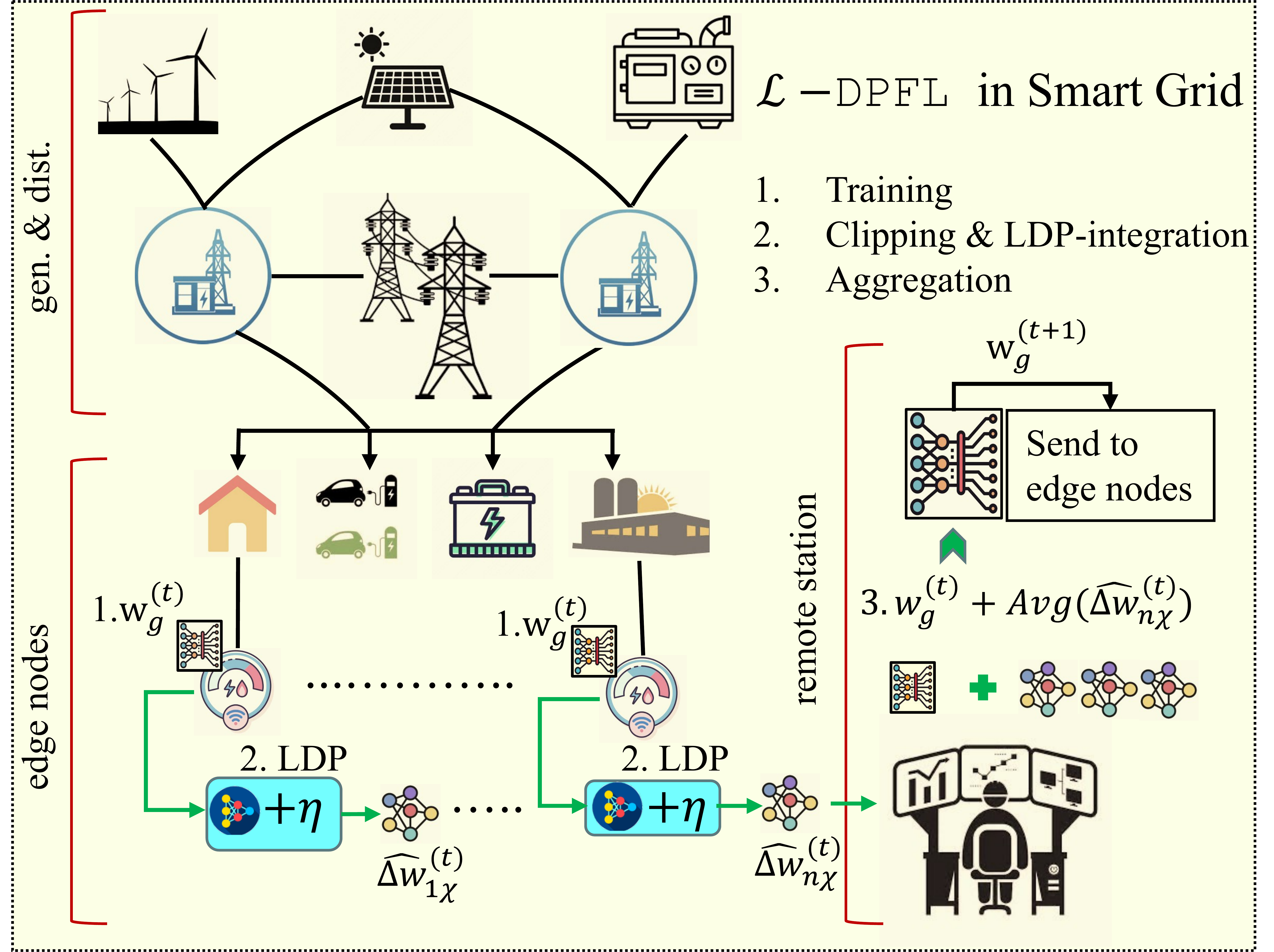}}
    \caption{{\tt $\mathcal{L}$-DPFL} architecture in smart grid metering network}
    \label{fig:FLInSG}
\end{figure}

The {\tt $\mathcal{L}$-DPFL} framework we use to model our proposed attack is similar to the smart metering network of Fig. \ref{fig:FLInSG}. In practice, a multi-layer network is more common and realistic as pointed out by \cite{zhou2022differentially}. Nonetheless, in a multi-layer FL network, the aggregation may happen in multiple layers but the FL training and LDP integration are carried out only at the edge layer. Even if the DP-noise is added into the successive layers to realize more privacy, that might introduce a new attack vector and opportunities for our strategic attacker. Therefore, for simplicity but without loss of generality, we consider a two-layer network where (i) \textit{edge nodes} are acting as FL clients and (ii) \textit{remote station} as FL server or aggregator. Our overall {\tt $\mathcal{L}$-DPFL} process can be described as follows.

\textbf{Training phase.} Consider that {\tt $\mathcal{L}$-DPFL} method consists of one \textit{remote station} and $n$ randomly selected edge nodes out of total $\mathcal{K}$ available edge nodes. The edge nodes are assumed to have their neural networks but with similar structures. At FL episode $t$ ($t \in \{1,2,...,T\}$), the \textit{remote station} distributes the global model parameters $w^{(t)}_g \in \mathbb{R}$ to $n$ edge nodes. The edge nodes initialize their network with those model parameters. Then they perform local optimization. Particularly, $k$th edge node randomly samples a dataset $\mathpzc{J}^{(t)}_k$ from its entire local training dataset $\mathpzc{D}_k$ and 
perform several steps of mini batch gradient descent to obtain the trained local model parameters $w^{(t)}_k$. Next, it computes the local model updates as $\Delta w^{(t)}_k = w^{(t)}_k - w^{(t)}_g$.

\textbf{Update clipping phase.}
Let $w$ is a weight vector, i.e., $w = (w_1, w_2, ..., w_d)$ and $\lVert p \rVert$ denotes the $\ell_2$-norm of a $q$-dimensional vector $p = (p_1, p_2, ..., p_q)$, i.e., $\lVert p \rVert = \sqrt{\sum_{i=1}^{q}p^2_i}$. Assume that $\mathpzc{W}$ is the maximum $\ell_2$-norm value of all weights for any given weight vector $w^{(t)}_k$ and sampled dataset $\mathpzc{J}^{(t)}_k$, i.e., $\mathpzc{W} = max_{w^{(t)}_{k} \in \mathbb{R}, \mathpzc{J}^{(t)}_k \in \mathpzc{D}_k}\mathbb{E}\left[\lVert w^{(t)}_k(\mathpzc{J}^{(t)}_k) \rVert \right]$. To keep the model usable and prevent over-fitting, each edge node clips their local model updates by a clipping threshold value $\mathcal{C} \in (0,\mathpzc{W}]$ as 
\begin{equation}
    \Delta w^{(t)}_{k\chi} \stackrel{\text{clip}}{=} \Delta w^{(t)}_k/max(1, \frac{\lVert \Delta w^{(t)}_k \rVert}{\mathcal{C}})
    \label{eqn:clipping}
\end{equation}
where $\chi$ denotes the clipping technique.

\textbf{LDP-integration phase.}
After clipping the local model updates by clipping threshold $\mathcal{C}$, the $k$th node implements the $(\varepsilon,\delta)$-LDP by adding a Gaussian noise component $\eta_k$. Since,  $\Delta w^{(t)}_{k\chi}$ is bounded by $\mathcal{C}$ and can be changed at most by $\mathcal{C}$, the local sensitivity, $\mathcal{S}$ of the aggregation operation is equivalent to $\mathcal{C}$. Therefore, the Gaussian noise variance of each dimension is proportional to $\mathcal{S}^2$, i.e., $\eta_k \sim \mathcal{N}(0,\mathcal{S}^2 \sigma^2_k \mathbb{I}_q)$ for some $\sigma^2_k > 0$, where $\mathbb{I}_q$ is the $q \times q$ identity matrix. Then, the noisy clipped local model updates can be represented as 
\begin{equation}
    \widetilde{\Delta w}^{(t)}_{k\chi} = \Delta w^{(t)}_{k\chi} + \eta_k \sim \mathcal{N}(0, \mathcal{S}^2\sigma^2_k \mathbb{I}_q)
    \label{eqn:noisyClipped}
\end{equation}
The noisy clipped local model updates from all edge nodes $\xi^{(t)} \leftarrow \{\widetilde{\Delta w}^{(t)}_{k\chi}\}_{k=1}^n$ are sent to the server for aggregation.

\textbf{Aggregation phase.}
At the \textit{remote station}, the noisy clipped local model updates are aggregated to obtain the new global model. Formally, the new global model can be expressed as
\begin{equation}
    w^{(t+1)}_g = w^{(t)}_g + \frac{1}{n} \sum_{k=1}^{n} \widetilde{\Delta w}^{(t)}_{k\chi}
    \label{eqn:newModel}
\end{equation}
The mechanism of {\tt $\mathcal{L}$-DPFL} is pseudocoded in algorithm \ref{algo1}.
\RestyleAlgo{ruled} 
\begin{algorithm}[!t]
\SetKwInput{KwNotation}{Notation}
\SetKwInput{KwInput}{Input}
\SetKwInput{KwOutput}{Output}  
\SetKwInput{KwPrivacy}{Privacy Guarantee}  
\DontPrintSemicolon

\KwInput{$N$, $\sigma$,  $\mathcal{C}$, $T$, $\alpha$}
  \KwOutput{New global model parameters, $w^{(t)}_g$}
  \KwData{Mini batch of training set, $\{\mathpzc{J}_k \subset \mathpzc{D}_k\}_{k=1}^n$}
  \KwPrivacy{satisfies ($\varepsilon, \delta$)-{\tt LDP} with Gaussian noise $\mathcal{N}(0, \mathcal{S}^2\sigma^2_k \mathbb{I}_q)$}
  \SetKwFunction{NoisyUpdates}{{\tt NoisyUpdates}}

$w^{(t)}_g \leftarrow$ random initialization\\
Initialize privacy accountant, $\Pi (\varepsilon, \mathcal{K})$\\
\For{each $t = 1,2,..., T$ episode}
{
$\delta \leftarrow \Pi(n_t, \sigma_t)$\\
\lIf{ $\delta >\tau_{\delta}$}{return $w^{(t)}_g$}
\lElse{
$\xi^{(t)} \leftarrow $ {\tt NoisyUpdates}($N,\sigma,C,w^{(t)}_g$)
$w^{(t+1)}_g = w^{(t)}_g + \frac{1}{n} \sum_{k=1}^{n} \xi^{(t)}$
}}

 \SetKwProg{Fn}{Function}{:}{}
  \Fn{\NoisyUpdates{$N,\sigma,C,w^{(t)}_g$}}{
  \For{each edge node $k \in N$}
  {$w^{(t)}_k \leftarrow w^{(t)}_g - \alpha . \frac{\partial \Phi(w^{(t)}_g,  \mathpzc{J}_k^{(t)})}{\partial w^{(t)}_g}$\;
  $\Delta w^{(t)}_k \leftarrow w^{(t)}_k - w^{(t)}_g$\\
  $\mathpzc{W} \leftarrow max_{w_{k} \in \mathbb{R}, \mathpzc{J}^{(t)}_k \in \mathpzc{D}_k}\mathbb{E}\left[\lVert w^{(t)}_k(\mathpzc{J}^{(t)}_k) \rVert \right]$\;
  Set clipping threshold $\mathcal{C}\in(0,\mathpzc{W}]$\;
  Clip the local model updates as
  $\Delta w^{(t)}_{k\chi} \stackrel{\text{clip}}{\leftarrow} \Delta w^{(t)}_k/max(1, \frac{\lVert \Delta w^{(t+1)}_k \rVert}{\mathcal{C}})$\; Add Gaussian noise to obtain
  $\widetilde{\Delta w}^{(t)}_{k\chi} \leftarrow \Delta w^{(t)}_{k\chi} + \eta_k \sim \mathcal{N}(0, \mathcal{S}^2\sigma^2_k \mathbb{I}_q)$
  }
  Set $\xi^{(t)} \leftarrow \{\widetilde{\Delta w}^{(t)}_{k\chi}\}_{k=1}^n$\;
        \KwRet $\xi^{(t)}$\;
        
  }

\caption{{\tt $\mathcal{L}$-DPFL} Protocol. $N$: Set of edge nodes with cardinality $\mathcal{K}$, $\sigma^2$: variance, $\mathcal{C}$: Clipping param., $T$: Total episode, $\alpha$: Learning rate, $\varepsilon$: Privacy loss, $\delta$: Privacy leakage probability, $\mathcal{S}$: Sensitivity, $\mathpzc{D}$: Training dataset, $\mathpzc{W}$: Max $\ell_2$-norm, $\xi$: Noisy clipped local model updates, $\eta$: Gaussian noise, $\mathcal{N}$: Noise profile, $\Pi$: Privacy accountant, $w$: Model parameter}
\label{algo1}
\end{algorithm}
\subsection{{\tt $\alpha$-MPELM}: An Adaptive Model Poisoning through Episodic Loss Memorization Technique}
\label{adaptiveAttack}
Following the adversarial update procedure, as described in section \ref{AdversarialUpdate}, we assume the total $m$ compromised nodes in the above {\tt $\mathcal{L}$-DPFL} environment. For a satisfactory attack impact $\mu_a^* = \theta + \sqrt{2\gamma}\sigma_x$ and stealthiness, the adversary tunes $\gamma$ at each FL episode and draws false noise $\eta_a \sim \mathcal{N}_a(\mu_a, \frac{\mathcal{S}}{\varepsilon})$ from the adversarial distribution $f_a^*$. For finding the \textit{episodic degree of poisoning, $\gamma^{(t)}$} at every $t$ episode, the attacker follows our adaptive model poisoning process through episodic loss memorization ({\tt $\alpha$-MPELM}) technique which is as follows.

\textbf{Choosing the initial degree of poisoning.}
Usually, the losses remain high in the first few episodes of FL for all local models. However, they gradually decrease to achieve convergence with the learning progression. Therefore, it is harder for the anomaly detector to distinguish benign models from the truly malicious models in the first few episodes than in the last ones. Particularly, the anomaly detector could- (1) remove all (benign and malicious) local models that are beyond the detection threshold or, (2) be undecided, and allow all local models for the first few episodes. Considering both cases, the initial degree of poisoning $\gamma_0$ should be chosen as close as possible to $\varepsilon$. It thereby ensures that the gradients do not explode in the upcoming episodes but also does not trim too much of the adversarial contribution.

\textbf{Calculating episodic loss.} At the beginning of any particular episode $t$, the attacker measures the validation losses of the updated global model (just received from the \textit{remote station}) for all compromised nodes $m$. Then, the attacker takes an average of those validation losses. Particularly, at episode $t$, for $i$th compromised node having a local validation dataset $\mathpzc{V}^{(t)}_i$ and received global model update $w^{(t)}_g$,  $i$'s validation loss $\mathcal{L}^{(t)}_i$ is computed through a loss function $\ell(w^{(t)}_g, \mathpzc{V}^{(t)}_i)$. Therefore, the average validation loss for all $m$ nodes at episode $t$ is
\begin{equation}
    \widetilde{\mathcal{L}}^{(t)}_{m} = \frac{1}{m}\sum_{i \in M, i = 1}^{m} \ell(w^{(t)}_g, \mathpzc{V}^{(t)}_i)
    \label{eqn:avgValidLoss}
\end{equation}
where $M$ is the set of compromised nodes. Then the attacker computes a loss ratio before appending $ \widetilde{\mathcal{L}}^{(t)}_{m}$ into the list of all episodic losses, ($[\widetilde{\mathcal{L}}^{(t)}_{m}]_{t=1}^{t}$). 

\textbf{Computing loss ratio.}
The loss ratio is computed between the average of current validation loss, $\widetilde{\mathcal{L}}^{(t)}_{m}$ and the average of all previous validation losses, $\widetilde{\mathcal{L}}^{(-t)}_{m}$. We represent all previous FL episodes before episode $t$ as $(-t)$. Formally, the loss ratio, $\mathpzc{R}$ can be expressed as follows.
\begin{equation}
    \mathpzc{R} = \widetilde{\mathcal{L}}^{(t)}_{m} \bigg/ \widetilde{\mathcal{L}}^{(-t)}_{m} \;\;\;\; \forall \; \widetilde{\mathcal{L}}^{(-t)}_{m} \neq 0; \; t>1
    \label{eqn:LossRatio}
\end{equation}
Since this process utilizes the validation losses of the previous episodes (or, episodes) to compute the loss ratio, $\mathpzc{R}$, we name it as \textit{episodic loss memorization} process. 

\textbf{Updating episodic degree of poisoning, $\boldsymbol{\gamma^{(t)}}$-value.}
Loss ratio $\mathpzc{R}>>1$ implies that the gradients of the current global model parameters have taken a turn towards the inverse of the direction along which the gradients of previous global models have descended (i.e., jumping out of the global minimum valley). This could happen due to several reasons including but not limited to the high learning rate, DP or external noise, etc. Adding more noise into this \textit{runaway gradient ascent} would only incur more losses in the following episodes which in turn, may lead to easier attack detection. To overcome this, the attacker can stop the model poisoning for that particular episode $t$ (i.e., $\gamma^{(t)} = 0$) and start again when $\mathpzc{R}\approx1$. Nonetheless, in other cases, poisoning could be stopped only partially (i.e., $\gamma^{(t)} \approx 0$) for a subset of the compromised nodes to experience similar outcomes.

On the other hand, in the case of $\mathpzc{R}<<1$, the attacker needs to increase the attack impact to maintain the attack persistence. One way to effectively achieve that is to adjust $\gamma^{(t)}$ in proportion to the loss ratio since the loss ratio reflects the most recent states of the entire FL process. i.e., $\gamma^{(t)} = \gamma + \rho \cdot \mathpzc{R}\cdot\gamma$ where $\rho$ is a factor of proportionality and $\gamma$ is most recent non-zero degree of poisoning. This ensures that the final global model has a substantial test loss and becomes sub-optimal (optimal damage). For the rest of the cases (i.e., $\mathpzc{R} \approx 1$), $\gamma^{(t)}$ is reduced as $\gamma^{(t)} = \gamma - \rho \cdot \mathpzc{R}\cdot\gamma$ to ensure the malicious updates do not deviate too much from the benign updates (avoiding detection) in the subsequent episodes. Then, $\gamma$ is updated with the most recent non-zero value of $\gamma^{(t)}$ to serve the next episode. Finally, the current average validation loss $\widetilde{\mathcal{L}}^{(t)}_{m}$ is appended in the episodic loss list. 
The pseudocode of {\tt $\alpha$-MPELM} is given in algorithm \ref{algo2}.

\RestyleAlgo{ruled} 
\begin{algorithm}[!t]
\SetKwInput{KwNotation}{Notation}
\SetKwInput{KwInput}{Input}
\SetKwInput{KwOutput}{Output}  
\SetKwInput{KwPrivacy}{Privacy Guarantee}  
\DontPrintSemicolon

  \KwInput{$w^{(t)}_g$}
  \KwOutput{$\gamma^{(t)}$}
  \KwData{$\{\mathpzc{V}^{(t)}_i\}_{i=1}^m$ $\;\forall$ $\mathpzc{V}^{(t)}_i \neq \mathpzc{J}^{(t)}_i$}
initialize: $\gamma \leftarrow \gamma_0$ where $\gamma_0 \approx \varepsilon$\\
\For{each $t = 1, 2,..., T$ episode}
{
set: $\mathcal{L}^{(t)}_{m} \leftarrow 0; \mathpzc{R} \leftarrow 0$\\
\For{each compromised node $i = 1, 2, ..., m$}{
measure: $\mathcal{L}^{(t)}_{i} \leftarrow  \ell(w^{(t)}_g, \mathpzc{V}^{(t)}_i)$\\
calculate: $\mathcal{L}^{(t)}_{m} = \mathcal{L}^{(t)}_{m} + \mathcal{L}^{(t)}_{i}$
}
current avg. loss: $\widetilde{\mathcal{L}}^{(t)}_{m} = \frac{1}{m}(\mathcal{L}^{(t)}_{m})$\\
avg. of episodic losses: $\widetilde{\mathcal{L}}^{(-t)}_{m} = Avg([\widetilde{\mathcal{L}}^{(e)}_{m}]_{e=1}^{(t-1)})$\\
Loss ratio: $\mathpzc{R} = \widetilde{\mathcal{L}}^{(t)}_{m} \bigg/ \widetilde{\mathcal{L}}^{(-t)}_{m} \;\;\;\; \forall \; \widetilde{\mathcal{L}}^{(-t)}_{m} \neq 0; \; t>1$\\
episodic degree of poisoning: $\gamma^{(t)} =
\begin{cases}
0, & \text{if}\; \mathpzc{R} >> 1 \\
\gamma + \rho\cdot\mathpzc{R}\cdot\gamma, & \text{if}\; \mathpzc{R} << 1\\
\gamma - \rho\cdot\mathpzc{R}\cdot\gamma, & \text{otherwise}
\end{cases}$\\
Then, save $\gamma$ for the next episode as follows: \\\lIfElse{$\gamma^{(t)} \neq 0$}{$\gamma \leftarrow \gamma^{(t)}$}{$\gamma$}
call: sub-processes to inject false noise with $\gamma^{(t)}$\;
append: $\widetilde{\mathcal{L}}^{(t)}_{m}$ into $[\widetilde{\mathcal{L}}^{(t)}_{m}]_{t=1}^{(t-1)}$ to obtain $[\widetilde{\mathcal{L}}^{(t)}_{m}]_{t=1}^{(t)}$
}

\caption{{\tt $\alpha$-MPELM} Technique. $\gamma$: degree of poisoning, $\gamma^{(t)}$: Episodic degree of poisoning, $\mathpzc{V}$: Validation dataset, $\mathcal{L}^{(t)}_m$: Current valid. loss, $\widetilde{\mathcal{L}}^{(t)}_m$: Current avg. valid. loss, $\widetilde{\mathcal{L}}^{(-t)}_m$: Previous avg. valid. loss, $\mathpzc{R}$: Loss ratio, $\rho$: proportionality factor}
\label{algo2}
\end{algorithm}

\section{Proposed RL-assisted Differential Privacy Level Selection ({\tt rDP}) Technique}
\label{sec:rDPmodelling}

From the optimal DP-exploited attack analysis as outlined in section \ref{sec:probFormulation} and \ref{sec:OurAttack}, it can be inferred that the increment of data privacy through DP-mechanism using large noise can potentially open a doorway for large poisoning attacks. One way to prevent that is to select a low DP level. However, low data privacy (i.e., a small amount of DP-noise) can then facilitate data privacy attacks. Therefore, an optimal value of privacy is desirable for any setting with DP.  We propose to achieve such an optimal privacy policy intelligently by tuning the \textit{privacy loss, $\varepsilon$} through reinforcement learning (RL) \cite{9469488}. We utilize the DP parameters
(privacy loss, information leakage probability, etc.) and historical federated loss to model the {\tt rDP} process.

\subsection{Defense Objectives}
\label{defenseObj}
The sole objective of the defender (or the designer of the CPCIs) is to \emph{design the learning process as fault-tolerant against the proposed attacks}. To achieve this, a proper understanding of the DP parameters and the threat model is necessary. At the same time, the designer needs to find out and set the optimal value of the privacy loss ($\varepsilon^*$) so that the attack surface is reduced and the attack impact ($\mu_a$) is minimized.

\subsection{The {\tt rDP} Algorithm}
\label{rdpAlgodescription}
The proposed {\tt rDP} process is pseudocoded in algorithm \ref{algo3}. We follow a learning approach based on Q-learning. The Q-learning follows an action-value function that gives the expected utility of taking
a given action in a given state.
\subsubsection{State space}
\label{stateSpace} We assume that the \textit{state} is initialized as soon the learning starts. We define the \textit{state space} as, $S = (m_l, f_l, \varepsilon)$ where $m_l$ represents the set of attacker's loss, $f_l$ denotes the historical federated loss, and $\varepsilon$ is the set of privacy loss. 
For the design purpose of the {\tt rDP} process, the set of attacker's losses ($m_l$) can be computed through several experiments in advance following the attack methodology as outlined in section \ref{adaptiveAttack}. Moreover, to obtain a realistic set of attacker's loss, the experiments should be conducted for multiple values of the episodic degree of poisoning ($\gamma^{(t)}$). On the other hand, the federated loss ($f_l$) can be simply measured by validating the global federated model considering a non-adversarial configuration. However, for the integrity and the accuracy of the {\tt rDP} process, both $m_l$ and $f_l$ need to be measured using the identical values of $\varepsilon$ from the loss set.

\subsubsection{Action space}
\label{actioneSpace}
We consider the event-driven manner approach where the defensive agent makes a decision when a new event occurs. The agent observes the federated environment's current state, $s \in S$ for making one of the decisions as described in the action space, $\mathcal{A}$. The \textit{action space} is defined as, $\mathcal{A} = \{increase,\; decrease,\; static\}$. To fine-grain the agent's action-making process, we consider that the agent can increase or decrease privacy loss, $\varepsilon$ by a single unit or double unit at any state, $s \in S$ and take respective action, $i \in \mathcal{A}$. 

\subsubsection{Reinforcement reward}
\label{reward}
In RL, the reward function motivates the defensive agent to decide on the learning objective. In each episode, the reward signal changes depending on the received input data. For defense against the proposed attack, \textit{the objective for the agent is to minimize the maximum attack accuracy as well as maximize the federated accuracy}. We assume that the maximum and minimum thresholds are set and regulated by the {\tt $\mathcal{L}$-DPFL} system designer. The \textit{reward function} is defined by the following equation as in (\ref{eqn:reward}),
\begin{equation}
    \begin{aligned}
    \mathbf{\beta}=\psi_1 \frac{m_l^{max}}{m_l} + \psi_2 \frac{f_l^{max}}{f_l} + \psi_3 \frac{1}{\varepsilon}
    \end{aligned}
    \label{eqn:reward}
\end{equation}
where $m_l^{max}$ and $f_l^{max}$ denotes the maximum value of the poisoning attack loss and the federated loss whereas $\psi_1$, $\psi_2$, and $\psi_3$ denotes the balancing parameters. Here, the exploration and exploitation dilemma is traded-off by the epsilon-greedy policy\cite{wunder2010classes}. We set the initial exploration probability at $1.0$, and gradually reduce the exploration probability over episodes until it matches the minimum exploration probability (which we assume $0.05$ in this paper). Moreover, for simplicity, we select the \textit{maximum number of episodes} as the stopping criterion or terminating condition.

\begin{algorithm}[!t]
\SetKwInput{KwInput}{Input}                
\SetKwInput{KwOutput}{Output}              
\DontPrintSemicolon
  
  \KwInput{$m_l, f_l, \varepsilon, S, \beta$}
  \KwOutput{Optimal privacy loss, $\varepsilon^* \leftarrow i$}

  \SetKwFunction{FrDP}{rDP}
 
  \SetKwProg{Fn}{Function}{:}{}
  \Fn{\FrDP{$m_l, f_l, \varepsilon$}}{
        \For{$\varepsilon_0$ in $\varepsilon$}{
        Set of States, $S_t = (m_l, f_l, \varepsilon_0)$\;
        Choose $i \in \mathcal{A}$ using epsilon-greedy policy \;
        Observe Reward, $r_{t+1}$ and State, $s_{t+1}$ \;
        Compute: $Q^{new}(s_t, i_t)$ $\leftarrow$ $(1-\alpha) \; . \;Q(s_t, i_t)$\;$\;\; + \; \alpha \; . \; [r_t + \zeta \; . \;  \substack{max\\i}\;Q(s_{t+1}, i)]$ \;
        Policy, $\pi(s) = \substack{arg\;max\\\pi}\; Q^{*}(s,i)$
        }
        \KwRet $i \leftarrow \pi^* (s)$\;
  }
\caption{{\tt rDP} process. $m_l$: Attacker's loss, $f_l$: Federated loss, $\varepsilon$: Privacy loss set, $S$: State set, $\mathcal{A}$: Action set, $\beta$: Reward func., $r$: Reward, $\alpha$: Learning rate, $Q$: Q-table, $i$: Action, $s$: State, $\pi$: Policy}
\label{algo3}

\end{algorithm}

\subsection{Convergence Analysis of {\tt rDP} Process}
\label{convergenceAnalysis}
The convergence of the proposed algorithm can be evaluated based on the average values of the $\Delta Q(s,i)$ of all $\Delta Q = Q^{new}(s_t, i_t) - Q(s_t, i_t)$ where state, $s \in S$ and action, $i \in \mathcal{A}$. The idea is to show the $Q$-value of the proposed {\tt rDP} process is converging to the optimal Q value ($Q^*$) defined by the Bellman equation in the stochastic case.
\begin{equation}
        Q^*(s,i) = r(s,i) + \zeta\; \substack{max\\i} \sum_{s_{t+1}} P(s_{t+1} | s, i)Q^*(s_{t+1},i) 
    \label{eqn:bellman}
\end{equation}
where $r(s,i)$ is the reward for taking the action $i \in \mathcal{A}$ giving the highest expected return, and $P(s_{t+1}|s, i)$ is the state transition probability. The expectation $E(Q^{new} (s_t, i_t))$ needs to converge to the optimal value $Q^*(s, i)$ as defined in (\ref{eqn:bellman}). However, without loss of generality, here we just show the deterministic case in which $Q^{new}(s_t, i_t)$ converges to $Q^*(s, i)$ defined as:
\begin{equation}
    \begin{aligned}
        Q^*(s,i) = r(s,i) + \zeta \; \substack{max\\i}\; Q^*(s_{t+1}, i)
    \end{aligned}
    \label{eqn:bellman2}
\end{equation}
Therefore, it can be stated that if the average of $\Delta Q (s, i)$ goes to zero, the proposed {\tt rDP} process is stable.
\begin{table}[!b]
\caption{Dataset description}
\centering
\resizebox{\columnwidth}{!}{%
\begin{tabular}{|l|l|}
\hline
\textbf{Dataset}                  & \textbf{Description} \\ \hline
Name                  & \begin{tabular}[c]{@{}l@{}}Individual Household \\ Electric Power Consumption\end{tabular} \\ \hline
Number of measurement    & 2,075,259   \\ \hline
Data collection range &  \begin{tabular}[c]{@{}l@{}}Dec 2006-Nov 2010\\ ($\sim$47 months)\end{tabular}  \\ \hline
Data missing percentage  & 1.25\%      \\ \hline
Data recording frequency & per minute  \\ \hline
\end{tabular}%
}
\label{table:datasetTable}
\end{table}

\begin{table}[!ht]
\caption{Hyperparameters}
\adjustbox{max width=\columnwidth}{%
\begin{tabular}{|l|c|c|c|}
\hline
\textbf{Parameters} & \textbf{Values} & \textbf{Parameters} & \textbf{Values} \\ \hline

Optimizer       & Adamax            & $\mathcal{K}$     & \{100, 1000, 10000\} \\ \hline
Loss metric     & MSE               & $n$               & \{30, 300, 3000\} \\ \hline
Hidden layers   & 2                 & $m$               & \{2, 7, 70\}  \\ \hline
Batch size      & 32                & $\varepsilon$     & \{0.5, 0.7, 1.0\}  \\ \hline
Valid. size     & 20\%              & $\delta$          & 0.001 \\ \hline
Activation      & ReLU             & $\gamma$          & \{1, 2, 3\}  \\ \hline
Early stop      & Enabled           & $\alpha$          & 0.001  \\ \hline

\end{tabular}
}
\label{table:hyperparameters}
\end{table}

\begin{figure*}[!t]
    \centerline{\includegraphics[width=1.25\linewidth]{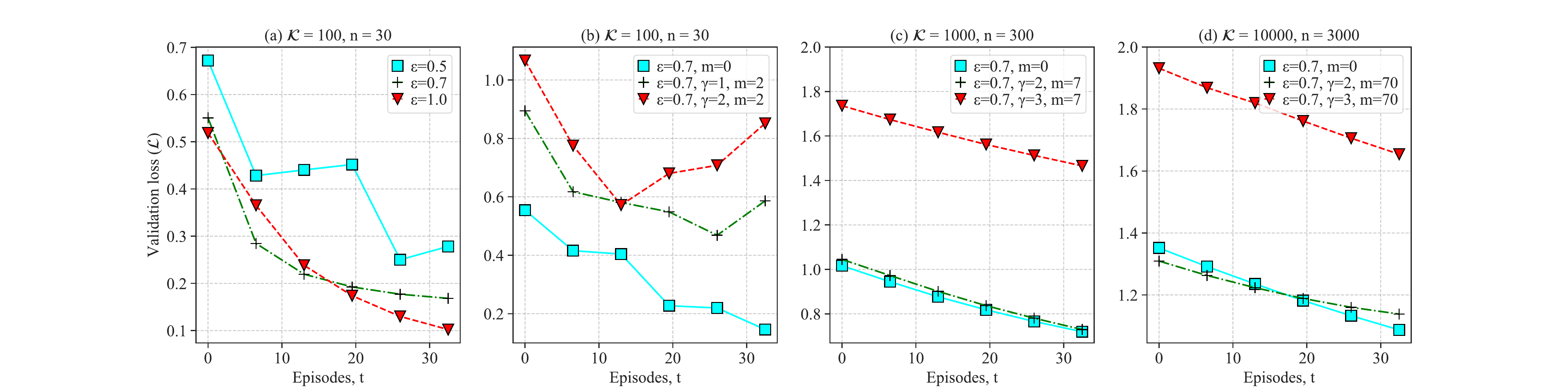}}
    \caption{Adversarial impact for varying privacy loss ($\varepsilon$), degree of poisoning ($\gamma$), and the number of malicious models ($m$)} 
    \label{fig:ImpactAnalysis}

\end{figure*}
\section{Experimental Analysis}
\label{sec:expAnalysis} 
In this part, we show how our proposed attack impacts the FL models over a smart grid example. Then, we evaluate our proposed attack to see if it can deceive the state-of-the-art anomaly detectors. We also evaluate our proposed defense policy for optimal convergence and attack detection.
\subsection{Dataset and Experimental Setup}
\label{datasetSetup}
To experimentally evaluate our proposed attack and defense policy, we use a smart grid dataset (Individual household electric power consumption dataset \cite{hebrail2012individual}). Table \ref{table:datasetTable} enlist some of the important features of the dataset. 
Although the dataset contains $1.25$\% missing records, the size of the dataset ($2,075,259$ records) is sufficient for the practical demonstration purpose of our model. For {\tt $\mathcal{L}$-DPFL} environment, we select the parameters as stated in Table \ref{table:hyperparameters}.
We perform the experiments on a lambda tensorbook with 11th Gen Intel(R) Core(TM) i$7$-$11800$H @$2.30$GHz CPU, RTX $3080$ Max-Q GPU, $64$ GB RAM, $2$ TB storage, Windows $10$ pro ($64$-bit) OS, Python $3.9.7$, and PyTorch $1.10.0+$cpu.

\subsection{Adversarial Impact Analysis}
\label{AdversarialImpactAnalysis}
Fig. \ref{fig:ImpactAnalysis} illustrates the adversarial impacts when there is no anomaly detector in the system while Fig. \ref{fig:norm}-\ref{fig:mix} presents adversarial impacts when there is an anomaly detection technique.
From the result in Fig. \ref{fig:ImpactAnalysis}(a), we can infer that the validation loss ($\mathcal{L}$) increases if we increase the privacy level (i.e., decrease the privacy loss $\varepsilon$) even if there is no attacker. \textit{This supports the intuitive fact that the loss increases if we add more DP-noise to achieve more privacy}. Besides, the $\mathcal{L}$ further increases if there is an adversary, injecting malicious contribution to the model/s. For instance, in Fig. \ref{fig:ImpactAnalysis}(b), for the same number of clients and same $\varepsilon$, the global model containing only one malicious local model (i.e., $m = 1$) has more loss than fully benign ones ($m = 0$). \textit{Hence, it provides empirical evidence that the system performance further degrades (i.e., loss increases) if there is at least one malicious entity injecting adversarial DP-noise to make the global model sub-optimal}.
\subsubsection{Impact of the Degree of Poisoning}
\label{degreeOfPoisoning}
From Fig. \ref{fig:ImpactAnalysis}(c) and \ref{fig:ImpactAnalysis}(d), we can comprehend that the attack impact is small if $\gamma = 2$ compared to the cases where $\gamma = 3$. Particularly, $\mathcal{L}$ would remain almost the same as the non-adversarial configuration ($m=0$) if $\gamma$ remains very small. However, as the attacker continues increasing $\gamma$, the loss is kept growing and the model becomes sub-optimal. 

Nevertheless, if $\gamma$ is very high, the malicious models would contribute high losses to the global model, in which case, the anomaly detector would easily identify those as `anomalous'. Hence, the attacker tunes $\gamma$ and obtain $\gamma^{(t)}$ at every FL episodes following the {\tt $\alpha$-MPELM} process as presented in section \ref{adaptiveAttack}. We simulate this adaptive attack to find out if the attack can deceive the state-of-the-art anomaly detectors \cite{fang2020local,zhou2022differentially} in this context. Since the {\tt norm, accuracy}, and {\tt mix} detection technique of \cite{zhou2022differentially} have the same underlying operational mechanism as \textit{ERR, LFR}, and \textit{mix} methods of \cite{fang2020local}, we decide to evaluate our proposed attack model only with the latest ones (i.e., {\tt norm, accuracy}, and {\tt mix} detection of \cite{zhou2022differentially}). 

\begin{figure*}[!t]
    \centering
    \centerline{\includegraphics[width=2.1\columnwidth]{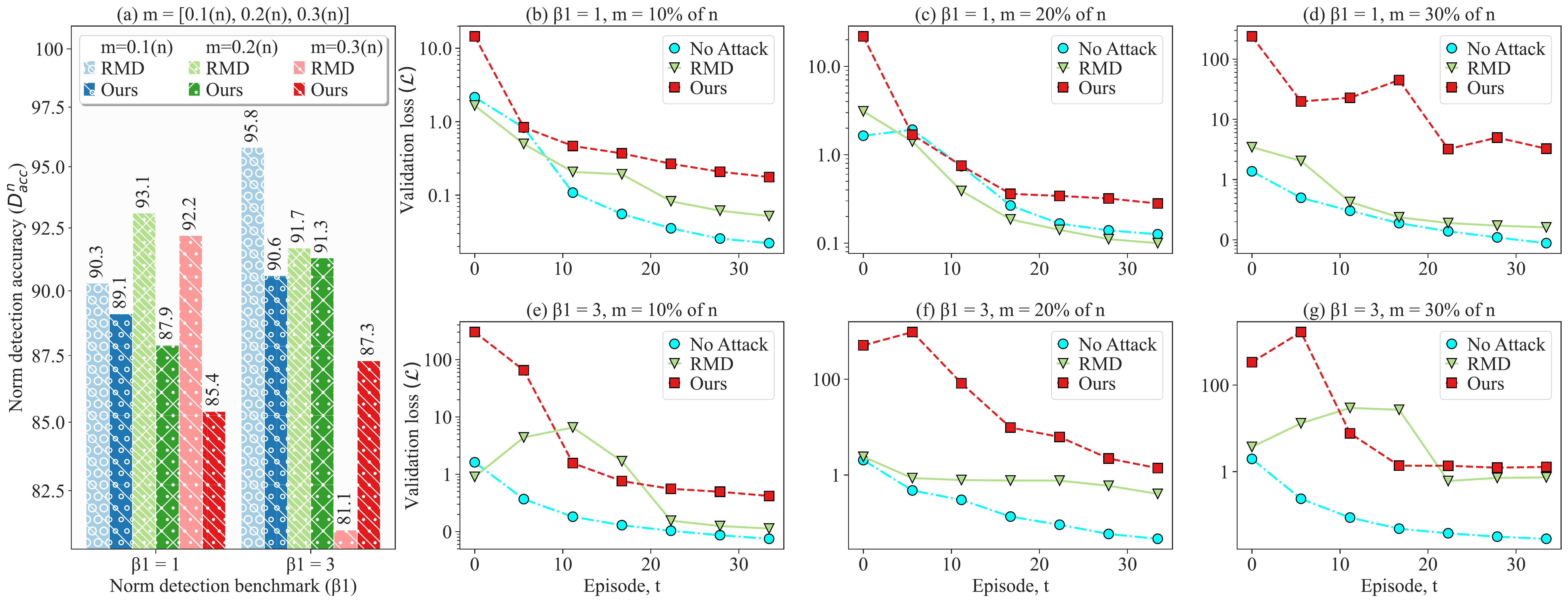}}
    \caption{Deceiving {\tt norm} detection: RMD attack vs our attack ($\varepsilon = 0.7$)}
    \label{fig:norm}
\end{figure*}
\textbf{Deceiving {\tt norm} detection.}
\label{deceivingNorm} According to the {\tt norm} detection, the aggregator computes a comparison standard for each local updates \cite{zhou2022differentially}. Particularly, it is calculated by taking the average of all the local model updates except that particular local model update itself. For instance, the comparison standard ($\Delta w^{(t)}_{i_{st}}$) of a particular noisy clipped local model update, $\widetilde{\Delta w}^{(t+1)}_{i\chi}$ is
\begin{equation}
    \Delta w^{(t)}_{i_{st}} = \frac{1}{n-1} (\sum_{k=1}^{n} \widetilde{\Delta w}^{(t)}_{k\chi} - \widetilde{\Delta w}^{(t)}_{i\chi})
    \label{eqn:normComparisonStd}
\end{equation}
Then, the square of the $L_2$ distance is computed as:
\begin{equation}
    d^{(t)}_i = \lVert \widetilde{\Delta w}^{(t)}_{i\chi} - \Delta w^{(t)}_{i_{st}} \rVert^2
    \label{eqn:normDist}
\end{equation}
Next, a reference value $e_1$ is calculated as:
\begin{equation}
    e_1 =
\begin{cases}
\frac{d^{(t)}_i}{\lVert \Delta w^{(t)}_{i_{st}} \rVert^2}, & \text{if}\; d^{(t)}_i < d_{max}\lVert \Delta w^{(t)}_{i_{st}} \rVert^2 \\
d_{max}, & \text{if}\; d^{(t)}_i \geq d_{max}\lVert \Delta w^{(t)}_{i_{st}} \rVert^2
\end{cases}
\label{eqn:normReference}
\end{equation}
where $d_{max}$ is the max. squared $L_2$ distance. Finally, the {\tt norm} detection accuracy of model $\widetilde{\Delta w}^{(t)}_{i\chi}$ is calculated as 
\begin{equation}
    rate^{norm}_i = 1 - max(0, e_1 - \beta_1)
    \label{eqn:normRate}
\end{equation}
where $\beta_1$ is a predefined {\tt norm} detection benchmark. If $rate^{norm}_i = 1$, the local update $\widetilde{\Delta w}^{(t)}_{i\chi}$ is flagged as non-anomalous whereas if $rate^{norm}_i < 1$, it is flagged as anomalous. The aggregator removes the detected anomalous models before aggregating the local models into a global model. To observe the {\tt norm} detection accuracy, \cite{zhou2022differentially} perform a random malicious device (RMD)-based attack where a group of malicious participants returns randomly generated local model parameters. However, the boundary of the RMD updates is not defined explicitly in their detection model. If the boundary is very large (i.e., RMDs return models with large parameters), the detection would be easy and the average {\tt norm} detection accuracy $D^{n}_{acc}$ may go up to $100\%$. On the other hand, if the boundary is small, the $D^{n}_{acc}$ would remain small. 
\begin{figure*}[!t]
    \centering
    \centerline{\includegraphics[width=2.1\columnwidth]{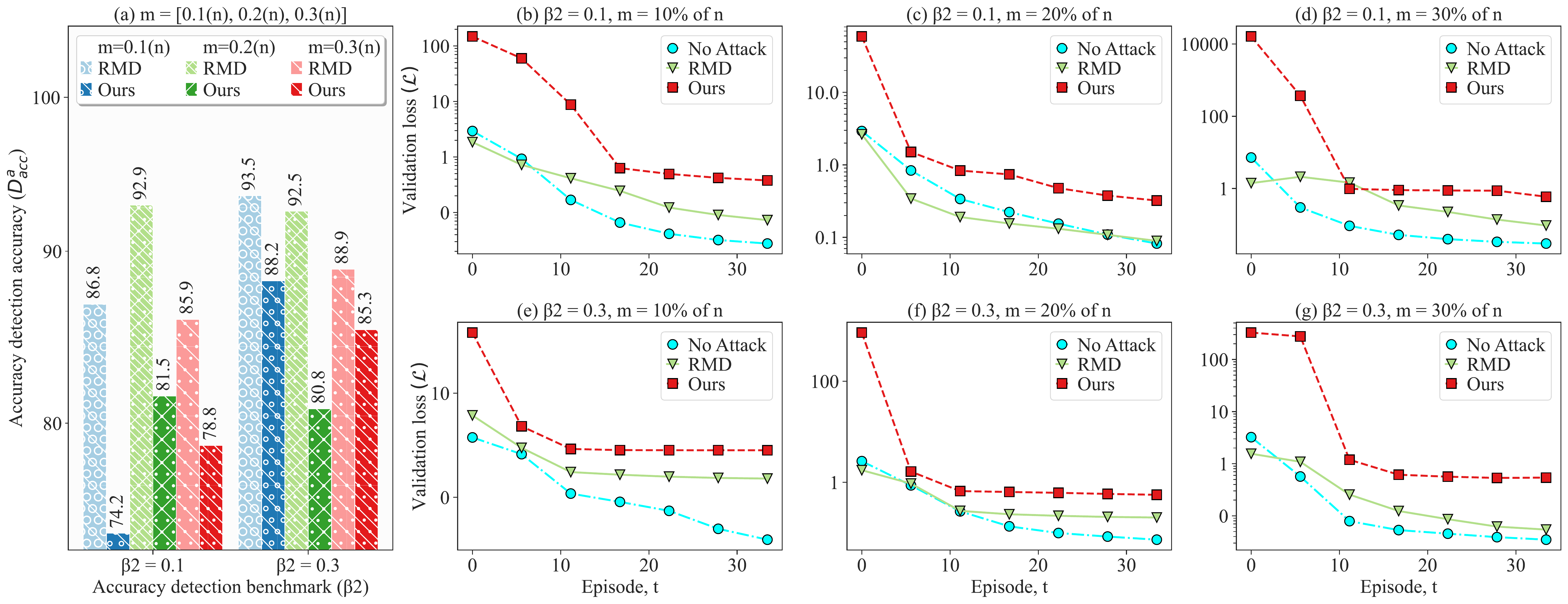}}
    \caption{Deceiving {\tt accuracy} detection: RMD attack vs our attack ($\varepsilon = 0.7$)}
    \label{fig:acc}
\end{figure*}
\begin{figure*}[!t]
    \centerline{\includegraphics[width=2.1\columnwidth]{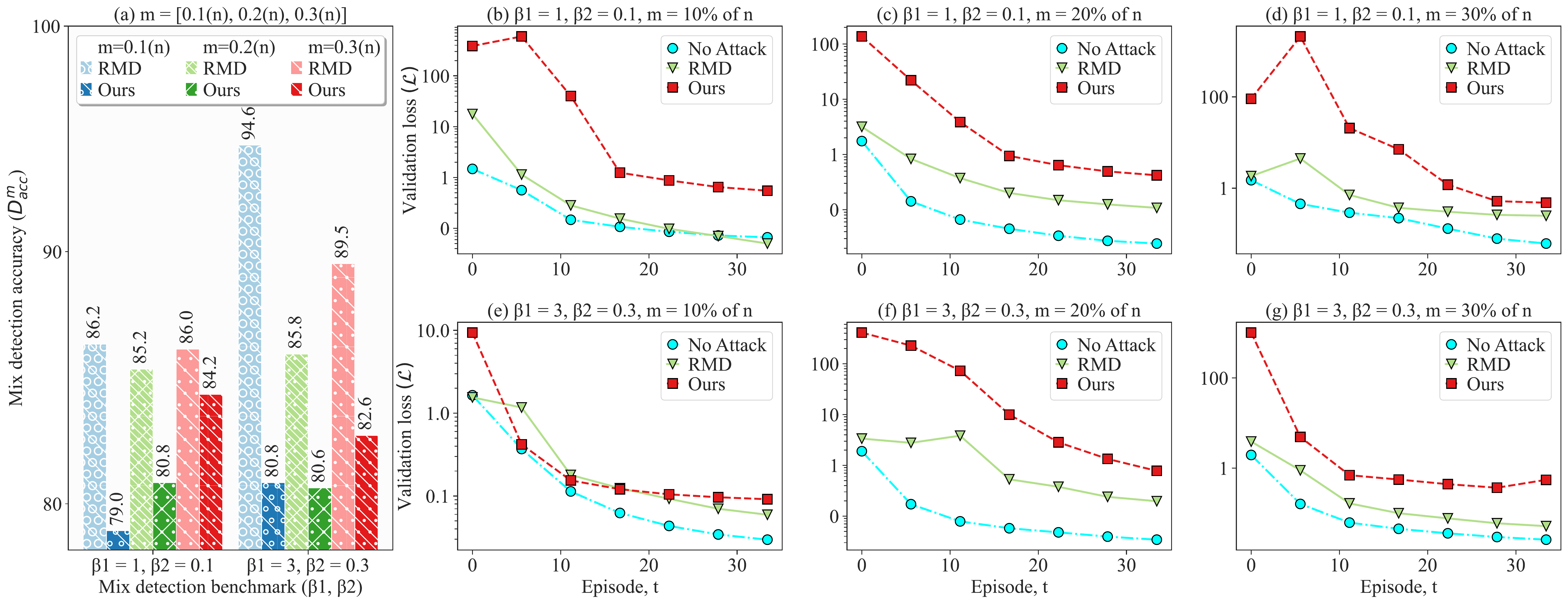}}
    \caption{Deceiving {\tt mix (norm+accuracy)} detection: RMD attack vs our attack ($\varepsilon = 0.7$)}
    \label{fig:mix}
\end{figure*}
To realize a practical RMD attack, we set this boundary equal to our clipping threshold, $\mathcal{C}$. 

From our experimental results as depicted in Fig. \ref{fig:norm}, we can see that $D^{n}_{acc}$ is decreasing due to our proposed attack with privacy loss, $\varepsilon = 0.7$. For instance, when $10\%$ of the total participants ($n$) are malicious (i.e., $m = 0.1(n)$) and $\beta_1 = 1$, $D^{n}_{acc}$ over RMD attack is $90.3$ whereas $D^{n}_{acc}$ over our proposed attack is $89.1$. The similar (or better) accuracy deviation is observed for $m = 0.2(n)$ and $m = 0.3(n)$ (illustrated and compared as the light and dark patches of each colors in Fig. \ref{fig:norm}(a)). Furthermore, the impact of our proposed attack reflects similar behavior when the {\tt norm} detection benchmark $\beta_1$ is further relaxed (i.e., $\beta_1 = 3$) except for one case when $m=0.3(n)$. For this particular case ($\beta_1 = 3, m = 0.3(n)$), our proposed attack incurs very large adversarial noise at the beginning, thus making it easier for the detector to identify and remove our anomalous models. However, as the learning progresses, the growth of the adversarial noise slows down. That is why the validation losses over our proposed attack remain larger than the RMD-attack (Fig. \ref{fig:norm}(g) red curve); even though our overall $D^{n}_{acc}$ is higher than the RMD-attack (the right-most red bar in Fig. \ref{fig:norm}(a)).   
When $D^n_{acc}$ is low, $\mathcal{L}$ is large. Fig. \ref{fig:norm}(b), (c) and (d) illustrates this loss increment for $10\%$, $20\%$, and $30\%$ malicious devices $n$ with $\beta_1 = 1$ respectively. Likewise, Fig. \ref{fig:norm}(e), (f) and (g) illustrates this loss increment for $10\%$, $20\%$, and $30\%$ malicious devices $n$ with $\beta_1 = 3$ respectively. In all cases, $\mathcal{L}$ is higher for our proposed attack than the `No Attack' and `RMD' attack scenarios.

\textbf{Deceiving {\tt accuracy} detection.}
\label{deceivingAcc}
Similar to the {\tt norm} detection methods, a comparison standard is calculated in {\tt accuracy} detection method following (\ref{eqn:normComparisonStd}) \cite{zhou2022differentially}. Now, instead of calculating the norm distance, an accuracy difference is calculated using a validation dataset. More specifically, two global models are computed using the local model update $\widetilde{\Delta w}^{(t)}_{i\chi}$ and its' comparison standard $\Delta w^{(t)}_{i_{st}}$. Then, \textit{accuracy tests} are conducted on the validation dataset for these two global models. However, since we are addressing a regression task in our experiment, it is more meaningful to conduct \textit{loss tests (mean squared error)} instead of the \textit{accuracy tests}. Therefore, if the \textit{loss test} results of  $\widetilde{\Delta w}^{(t)}_{i\chi}$ and $\Delta w^{(t)}_{i_{st}}$ are $\mathcal{L}^{(t)}_i$ and $\mathcal{L}^{(t)}_{st}$ respectively, then the loss difference is
\begin{equation}
    \Delta \mathcal{L}^{(t)}_i = 
    \begin{cases}
    0 & \text{if}\; \mathcal{L}^{(t)}_{st} \leq \mathcal{L}^{(t)}_i \\
    \frac{\mathcal{L}^{(t)}_{st} - \mathcal{L}^{(t)}_i }{\mathcal{L}^{(t)}_{st}} & \text{if}\; \mathcal{L}^{(t)}_{st} > \mathcal{L}^{(t)}_i
    \end{cases}
\end{equation}
Then, a reference value $e_2$ is calculated as
\begin{equation}
    e_2 = max\; (\Delta \mathcal{L}^{(t)})
\end{equation}
Finally, the {\tt accuracy} detection rate $rate^{acc}_i$ is computed as 
\begin{equation}
    rate^{accuracy}_i = 
    \begin{cases}
    1 - e_2 & \text{if}\; e2 > \beta_2\\
    1 & \text{if}\; e2 < \beta_2
    \end{cases}
\end{equation}
where $\beta_2$ is a predefined {\tt accuracy} detection benchmark. If $rate^{acc}_i = 1$, the local model update $\widetilde{\Delta w}^{(t)}_{i\chi}$ is flagged as non-anomalous whereas if $rate^{acc}_i < 1$, it is flagged as anomalous. 
To observe the accuracy of the {\tt accuracy} detection method, \cite{zhou2022differentially} perform a specialized malicious end device (SMD)-based attack where a group of malicious participants returns trained local model parameters so as to modify the sample label of a certain data category. But, as we mainly focus on the regression task with untargeted poisoning instead of the categorical classification task with targeted poisoning, we choose to conduct the same RMD attack instead of this targeted SMD attack. Similar to the previous, we set the boundary of the RMD attack equal to our clipping threshold to realize a practical and stricter attack. From Fig. \ref{fig:acc}(a), we can see that $D^{a}_{acc}$ is decreasing more due to our proposed attack (with privacy loss, $\gamma = 0.7$) than the RMD attack. At the same time, $\mathcal{L}$ is increasing due to the high misclassification of the local models as illustrated through Fig. \ref{fig:acc}(b)-\ref{fig:acc}(g).

\textbf{Deceiving {\tt mix (norm+accuracy)} detection.}
In the {\tt mix} detection, the {\tt norm} and {\tt accuracy} detection are combined. Specifically, the aggregator removes the local models that are detected by either {\tt norm} detection or {\tt accuracy} detection. The performance of the {\tt mix} detection over our proposed attack and the RMD attack is depicted in Fig. \ref{fig:mix}. We can observe the similar effects of our proposed attack on $D^{m}_{acc}$ as {\tt norm} and {\tt accuracy} detection.
\label{deceivingMix}
\begin{figure}[!t]
    \centerline{\includegraphics[width=0.8\columnwidth]{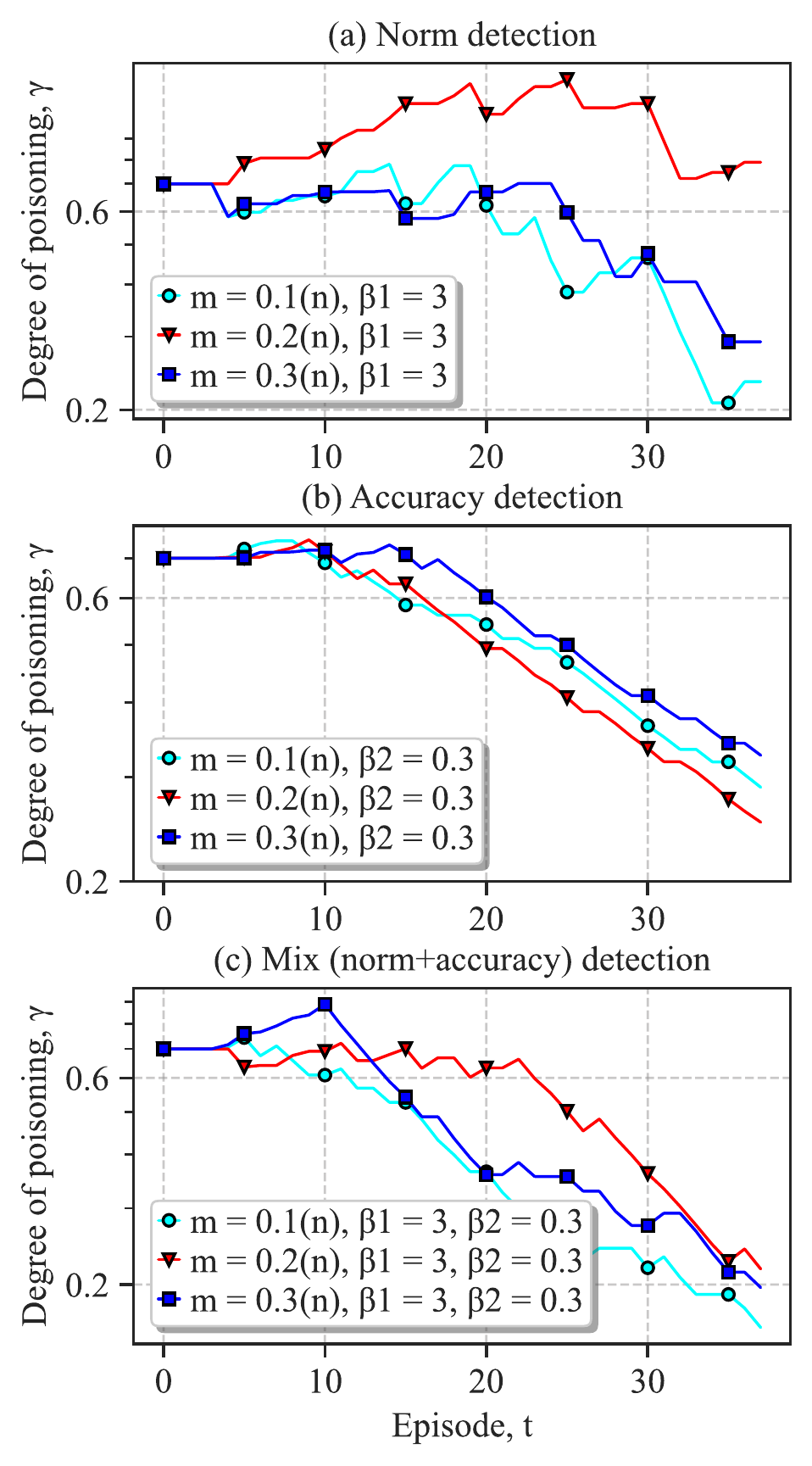}}
    \caption{Changes of $\gamma$ while deceiving the {\tt norm, accuracy}, and {\tt mix} detection. Similar results can be observed for $\beta_1 = 1$ and $\beta_2 = 0.1$}
    \label{fig:gamma}
\end{figure}

\textbf{Changes in the degree of poisoning.}
Fig. \ref{fig:gamma} illustrates the changes of $\gamma$ at every FL episodes while deceiving the {\tt norm, accuracy} and {\tt mix} detection techniques. As we can see from the figure, in most of the cases, $\gamma$ starts to decrease after a few episodes. However, following the adaptive {\tt $\alpha$-MPELM} process, when loss ratio $\mathpzc{R}>>1$, $\gamma^{(t)}=0$ (algorithm \ref{algo2}, line $11-13$). That is why in some cases $\gamma$ remains unchanged for consecutive rounds. For example, in Fig. \ref{fig:gamma}(a), $\gamma$ remains same for episodes $t = 21$ and $t = 22$.

\subsubsection{Impact of Attacker-Client Ratio}
\label{attackerClientRatio}
Another important criteria for successful (or unsuccessful) attack is the attacker-client ($m/n$) ratio. If the ratio is low, then the global model remains close to the optimal global model. However, if the ratio increases, the global model starts to deviate farther from the optimal model (if not diverge). This can be also observed in Fig. \ref{fig:norm}, Fig. \ref{fig:acc}, and Fig. \ref{fig:mix}. In most of the cases, the red lines keep deviating more and more from cyan lines if $m$ increases with respect to $n$.


\subsection{Performance Analysis of {\tt rDP}-technique}
\label{rDPperformance}
To defend the DP-exploited stealthy model poisoning attacks, the privacy level for the nodes needs to be chosen carefully during the design phase. Too much privacy leads to poor model performance and a wider poisoning window, whereas, very little privacy can expose the confidential and crucial operating information of the grid. To demonstrate the scenario, we perform several experiments for multiple values of the RL parameters. The results are presented in Table \ref{table:rDPAnalysisTable}. It can be inferred that for every discount factor, $\zeta$ (except $\zeta = 1.00$), at every level of learning rate ($\alpha$), $\Delta Q(s, i)$ value is almost zero when the learning ends. However, the average global reward, $R$ varies largely with the change of $\zeta$. The maximum reward is achieved when the learning rate, $\alpha = 0.001$, and the discount factor, $\zeta = 0.50$ ensures that the agent not only cares for the present reward but also considers the future reward equally.
\subsubsection{Reward Evaluation}
\label{rewardEval}
\begin{figure}[!t]
    \centerline{\includegraphics[width=0.8\columnwidth]{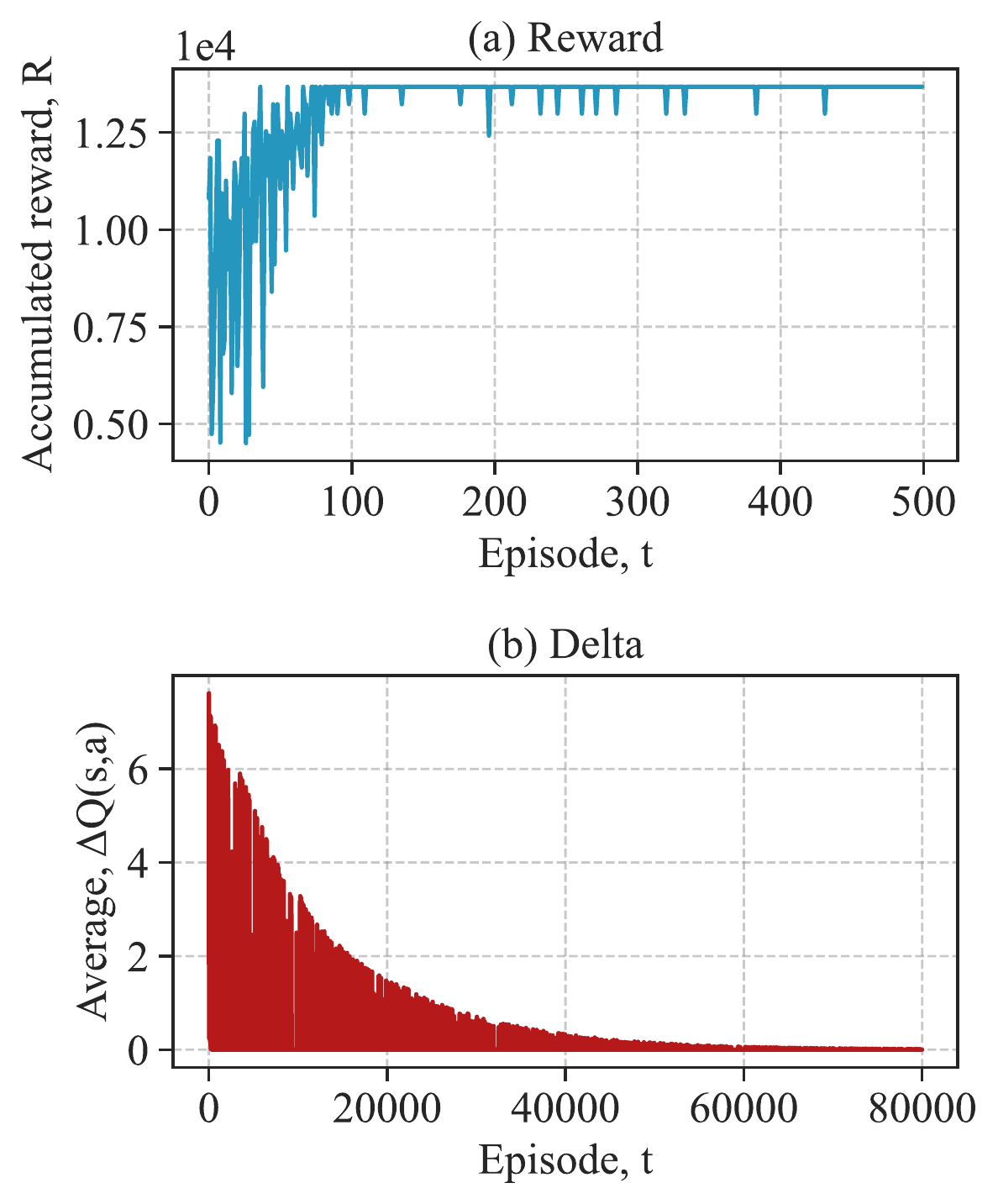}}
        \caption{(a) Accumulated rewards converge after a certain number of episodes (b) Convergence of $\Delta Q$-values.}
        \label{table:RLresult}
\end{figure}

The average global reward is computed as $R = \sum_{i=1}^{t}{r^i}$.
Fig. \ref{table:RLresult}(a) depicts the accumulated reward of the defender for discount factor ($\zeta = 0.50$) and learning rate ($\alpha = 0.001$). The RL agent learns optimal policy as the episodes increase. After sufficient episodes are executed, it converges into an optimal policy that ensures the desired privacy, utility, and security. From the result in Fig. \ref{table:RLresult}(a), we can see that the policy converges around episode $100$ and stays high ($\approx 13142.54$) for the rest of the period. That means the agent is making optimal actions at this stage.

\begin{table}[!t]
\caption{Performance analysis of the {\tt rDP} technique}
\centering
\resizebox{\columnwidth}{!}{%
\setlength{\extrarowheight}{0pt}
\addtolength{\extrarowheight}{\aboverulesep}
\addtolength{\extrarowheight}{\belowrulesep}
\setlength{\aboverulesep}{0pt}
\setlength{\belowrulesep}{0pt}
\arrayrulecolor{black}
\begin{tabular}{|c|c|r|r|} 
\toprule
\multirow{2}{*}{Learning rate, $\alpha$}                                                             & \multirow{2}{*}{Discount factor, $\zeta$}          & \multicolumn{2}{c|}{{\tt rDP} values}                                                                          \\ 
\cline{3-4}
                                                                                                                  &                                                                  & \multicolumn{1}{c|}{Reward, $R$}               & \multicolumn{1}{c|}{Delta, $\Delta Q(s,i)$}  \\ 
\hline
\multirow{4}{*}{$\alpha$ = 0.01}                                                                     & {}$\zeta$ = 1.00 & {}8475.33  & {} $4.13e−00$            \\ 
\hhline{|~---|}
                                                                                                                  & {}$\zeta$ = 0.50 & {}13019.58 & {}$2.86e−07$            \\ 
\hhline{|~---|}
                                                                                                                  & {}$\zeta$ = 0.20 & {}11726.13 & {}$4.25e−08$            \\ 
\hhline{|~---|}
                                                                                                                  & {}$\zeta$ = 0.15 & {}10368.59 & {}$3.32e−08$           \\ 
\hline
\multirow{4}{*}{$\alpha$ = 0.001}                                                                     & {}$\zeta$ = 1.00 & {}9718.91  & {} $2.92e−00$            \\ 
\hhline{|~---|}
                                                                                                                  & {}$\zeta$ = 0.50 & {}13142.54 & {}$6.24e−04$            \\ 
\hhline{|~---|}
                                                                                                                  & {}$\zeta$ = 0.20 & {}10190.45 & {}$1.63e−04$            \\ 
\hhline{|~---|}
                                                                                                                  & {}$\zeta$ = 0.15 & {}11249.85 & {}$2.94e−04$           \\ 
\hline
\multirow{4}{*}{$\alpha$ = 0.0001}                                                                   & {}$\zeta$ = 1.00 & {}9246.74  & {}$2.14e−01$            \\ 
\hhline{|~---|}
                                                                                                                  & {}$\zeta$ = 0.50 & {}11803.06 & {}$1.77e−04$            \\ 
\hhline{|~---|}
                                                                                                                  & {}$\zeta$ = 0.20 & {}10974.27 & {}$9.15e−05$            \\ 
\hhline{|~---|}
                                                                                                                  & {}$\zeta$ = 0.15 & {}11031.96 & {}$1.05e−04$            \\
\bottomrule
\end{tabular}}
\label{table:rDPAnalysisTable}
\end{table}

\subsubsection{Q-value Evaluation}
\label{qvalueEval}
As stated in the convergence analysis part (section \ref{convergenceAnalysis}), if the average of $\Delta Q(s,i)$ goes to zero, the proposed process is stable. From the result of Fig. \ref{table:RLresult}(b), we can see that for our proposed {\tt rDP} technique, the average of $\Delta Q(s,i)$ gradually goes to zero after $60,000$ episodes.
\subsubsection{Assisting Attack Detection}
\label{assistAttackDetection}
The reward function, $\beta$ as expressed in (\ref{eqn:reward}), takes care of the attacker's loss ($m_l$), federated loss ($f_l$), and  privacy loss ($\varepsilon$). The RL agent determines an action for each state, therefore, the standard value of the federated loss, $f_l^s$ can be calculated for that state. Now, if the observed federated loss for any particular state, $f_l^o$ does not match or differs significantly with $f_l^s$, the following cases can occur:

\begin{itemize}
    \item $f_l^s < f_l^o$: large-scale attacks are launched (high $\gamma$).
    \item $f_l^s \geq f_l^o$: either the system is not compromised or the degree of poisoning attack is sufficiently low that it can be neglected.
\end{itemize}
Therefore, the {\tt rDP} algorithm not only limits the attack surface by intelligently and automatically selecting the privacy loss level in the design phase but also facilitates the attack detection technique. 
Moreover, the privacy level selection through the proposed {\tt rDP}-technique need not be conducted only in real-time; rather it can be carried out offline through some test experiments during the design phase of the {\tt $\mathcal{L}$-DPFL} process. This ensures its applicability in limiting the attack surface for critical infrastructure operations.

\subsubsection{Limitations and Future Recommendations}
\label{limitations} A shortcoming of the {\tt DPFL} process is that it adds randomized noise in every episode for every client. Thus, the learning process can experience more fluctuations than in the non-DP environment. A reasonable solution to this problem could be bounding the model updates to a threshold.
However, too much clipping of the model updates may undermine the privacy protections of the DP. Hence, more research needs to be conducted to effectively clip the model updates without hampering privacy. Another limitation of this research could be the timing of the attack. For simplicity, we deliberately perform the attacks on the compromised models from the first episode. However, in practice, the attacks may start from any stage of the learning process. Intuitively, it should follow the same adversarial principle as ours since the other influencing parameters (e.g., privacy level, number of participants, etc.) remain the same.

We understand that several other advanced DP-FL algorithms are there in the literature and more are expected to be developed in the future. However, if these methods utilize the Gaussian noise to privatize the FL models, the attacker should have the opportunity to exploit that noise.  The anomaly detection-based defense techniques \cite{zhou2022differentially,fang2020local} focus on filtering the anomalous models. On the contrary, our rDP defense technique aims to limit the attack surface and disincentivize the attacker in the first place. One benefit of our defense method over anomaly detection-based defenses is that our defense serves in both the pre-attack (lowers attack surface) and the post-attack phases (degrades attack impact) whereas the others serve in only the post-attack phases (degrades attack impact through model filtering).

\section{Conclusion}
\label{sec:conclusionAndFutureWorks}
In this paper, we study the problem of model poisoning attacks in conjunction with FL and DP. Particularly, we find out an intelligent attacker can leverage the added Gaussian noise (to ensure DP) to perform a stealthy and persistent model poisoning attack in the FL domain. We show that our proposed {\tt $\mathcal{L}$-DPFL} attack degrades the accuracy of state-of-the-art detection techniques. As a countermeasure, we propose another novel defense strategy called {\tt rDP}. We show that the {\tt rDP} process converges to an optimal policy. To the best of our knowledge, our results are the first to consider a novel model poisoning threat in the context of {\tt DPFL}-driven CPCIs. We believe this study will open a new research area in the adversarial FL domain. In the future, we plan to extend this work toward the targeted model poisoning attacks using non-IID data.


\section*{Acknowledgement}
This work is supported by the Vingroup Joint Stock Company and supported by Vingroup Innovation Foundation (VINIF) under project code VINIF.2020.NCUD.DA094. The views, opinions, findings, and conclusions reflected in this publication are solely those of the authors and do not represent the official policy or position of the VINIF.

\bibliography{ref}
\nocite{*}
\bibliographystyle{IEEEtran}


%





\ifCLASSOPTIONcaptionsoff
  \newpage
\fi

%

\begin{IEEEbiography}[{\includegraphics[width=0.8in,height=0.8in,clip,keepaspectratio]{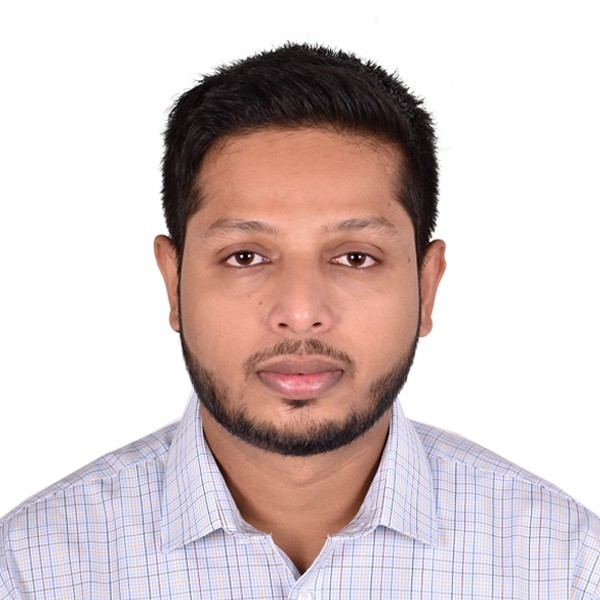}}]{Md Tamjid Hossain} is currently pursuing his Ph.D, in Computer Science and Engineering at the University of Nevada, Reno. He received his MS in Computer Science and Engineering from the University of Nevada, Reno, USA. His current research interest includes adversarial federated learning, differential privacy, and critical infrastructure security. Tamjid’s research outcomes have been published in top conferences including IEEE CNS, IEEE MSN, and so on.
\end{IEEEbiography}
\vspace{-50pt}
\begin{IEEEbiography}[{\includegraphics[width=0.8in,height=0.8in,clip,keepaspectratio]{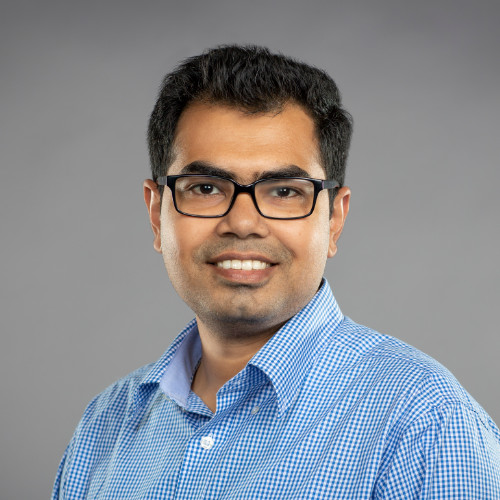}}]{Shahriar Badsha} is currently serving as a Senior Security Engineer at Bosch Engineering, North America. Before joining Bosch, he served as an assistant professor in cybersecurity in Computer Science and Engineering at the University of Nevada, Reno. He completed his Ph.D. in computer science and software engineering from RMIT University, Australia. He was also with data61, CSIRO, in Melbourne, Australia.
\end{IEEEbiography}
\vspace{-50pt}
\begin{IEEEbiography}[{\includegraphics[width=0.8in,height=0.8in,clip,keepaspectratio]{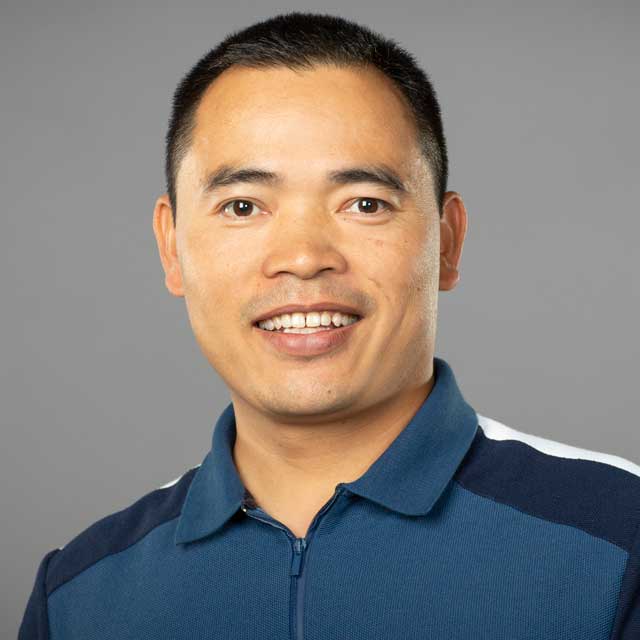}}]
{Hung (Jim) La} is an associate professor of Computer Science and Engineering at the University of Nevada, Reno. He is also the associate director of the INSPIRE Tier 1 University Transportation Center. He has authored over 116 research papers, and eight of his papers have won best conference paper awards and best paper finalists in the top-ranked robotics conferences (IROS 2019, SSRR 2018, ICRA 2017, ISARC 2015, etc.).
\end{IEEEbiography}
\vspace{-50pt}
\begin{IEEEbiography}[{\includegraphics[width=0.8in,height=0.8in,clip,keepaspectratio]{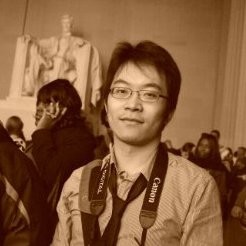}}]
{Haoting Shen} is an assistant professor at the School of Cyber and Technology, College of Computer Science and Technology, Zhejiang University. Earlier, he served as an assistant professor in the Dept. of Computer Science and Engineering at the University of Nevada, Reno. He completed his Ph.D. from Penn State University and postdoc from the University of Florida.
\end{IEEEbiography}
\vspace{-50pt}
\begin{IEEEbiography}[{\includegraphics[width=0.8in,height=0.8in,clip,keepaspectratio]{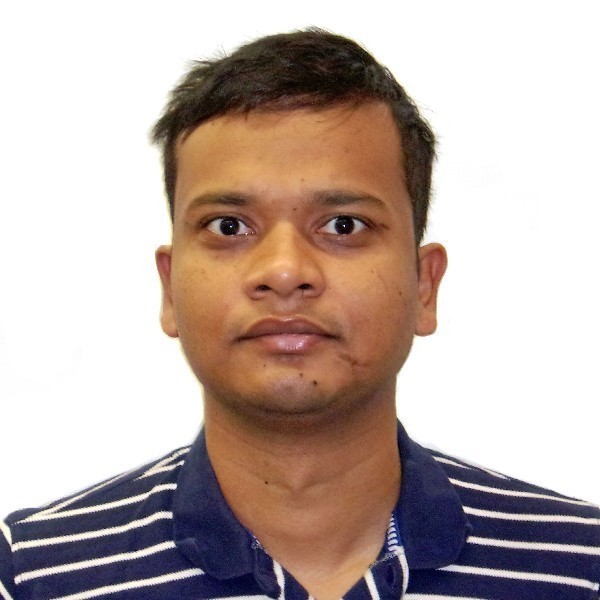}}]{Shafkat Islam} pursued an MS in Computer Science and Engineering at the University of Nevada, Reno. He is currently pursuing a Ph.D. at Purdue University, USA. His current research interests include responsible AI, cybersecurity, and connected autonomous vehicles. Shafkat's research outcomes have been published in top venues and high-impact journals and magazines including IEEE Network, IEEE IoT Journal, IEEE WoWMoM, IEEE LCN, and so on.
\end{IEEEbiography}
\vspace{-50pt}
\begin{IEEEbiography}[{\includegraphics[width=0.8in,height=0.8in,clip,keepaspectratio]{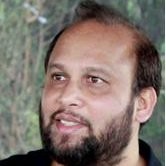}}]{Ibrahim Khalil} is a professor in Computer Science and Software Engineering, RMIT University, Melbourne, Australia. Ibrahim obtained his Ph.D. in 2003 from the University of Berne in Switzerland. He has several years of experience in Silicon Valley-based companies working on Large Network Provisioning and Management software. His research interests are in Privacy, Blockchain, network and data security, and secure data analysis including big data security.
\end{IEEEbiography}
\vspace{-40pt}
\begin{IEEEbiography}[{\includegraphics[width=0.8in,height=0.8in,clip,keepaspectratio]{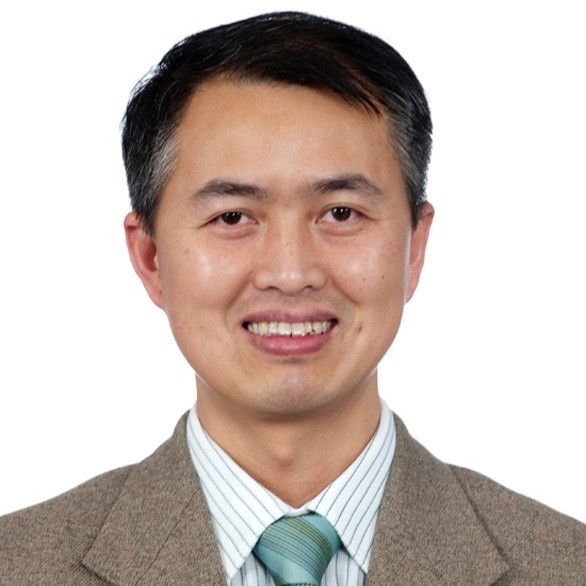}}]{Xun Yi} is currently a Professor of Computer Science and Software Engineering, RMIT University, Melbourne, Australia. He has published more than 200 research papers in international journals and conference proceedings. His research interests include applied cryptography, computer security, mobile, and wireless communication security, and data privacy protection. Prof. Yi has ever undertaken program committee members for more than 30 international conferences. From 2014 to 2018, he was an Associate Editor for IEEE Transactions on Dependable and Secure Computing.
\end{IEEEbiography}




\end{document}